\documentclass[preprint,review,10pt,3p]{elsarticle}

\usepackage{amssymb}

\usepackage{lineno}

\usepackage{lipsum}
\usepackage{amsmath} 
\usepackage{relsize} 
\usepackage{commath}
\usepackage{hyperref}
\usepackage{tabularx}
\usepackage{booktabs}
\usepackage{multicol}
\usepackage{makecell}
\usepackage{subfig}
\usepackage{cleveref}
\usepackage{xurl}
\usepackage{censor}
\usepackage{xcolor}
\newcommand{\hlcolor}{black} 
\renewcommand{\censor}{} 

\newcolumntype{Y}{>{\centering\arraybackslash}X}

\journal{Journal of Building Performance Simulation}

\date{October 21, 2024} 

\begin{document}

\begin{frontmatter}



\title{Rationalising data collection for supporting decision making in building energy systems using Value of Information analysis}

\author[EECi]{Max Langtry\corref{cor1}}\ead{mal84@cam.ac.uk}
\author[EECi,ATI]{Chaoqun Zhuang}
\author[EECi,ATI]{Rebecca Ward}
\author[EECi,Surrey]{Nikolas Makasis}
\author[EECi,Surrey]{Monika J. Kreitmair}
\author[EECi,ATI]{Zack Xuereb Conti}
\author[ATI,CSML]{Domenic Di Francesco}
\author[EECi,ATI]{Ruchi Choudhary}

\cortext[cor1]{Corresponding author}

\affiliation[EECi]{organization={Energy Efficient Cities Initiative, Department of Engineering, University of Cambridge},
            addressline={Trumpington Street},
            city={Cambridge},
            postcode={CB2 1PZ},
            country={UK}}

\affiliation[ATI]{organization={Data-Centric Engineering, The Alan Turing Institute},
            addressline={British Library},
            city={London},
            postcode={NW1 2DB},
            country={UK}}

\affiliation[Surrey]{organization={School of Sustainability, Civil \& Environmental Engineering, University of Surrey},
            city={Guilford},
            postcode={GU2 7XH},
            country={UK}}

\affiliation[CSML]{organization={Computational Statistics \& Machine Learning, Department of Engineering, University of Cambridge},
            city={Cambridge},
            postcode={CB3 0FA},
            country={UK}}

\begin{abstract}
The use of data collection to support decision making through the reduction of uncertainty is ubiquitous in the management, operation, and design of building energy systems. However, no existing studies in the building energy systems literature have quantified the economic benefits of data collection strategies to determine whether they are worth their cost. This work demonstrates that Value of Information analysis (VoI), a Bayesian Decision Analysis framework, provides a suitable methodology for quantifying the benefits of data collection. Three example decision problems in building energy systems are studied: air-source heat pump maintenance scheduling, ventilation scheduling for indoor air quality, and ground-source heat pump system design. Smart meters, occupancy monitoring systems, and ground thermal tests are shown to be economically beneficial for supporting these decisions respectively. It is proposed that further study of VoI in building energy systems would allow expenditure on data collection to be economised and prioritised, avoiding wastage.
\end{abstract}



\begin{keyword}
Data collection \sep Value of Information \sep Building energy management \sep Uncertainty quantification \sep Bayesian Decision Analysis


\end{keyword}

\end{frontmatter}



\section{Introduction} \label{sec:intro}


\subsection{Background} \label{sec:background}

Buildings account for 59\% of electricity consumption and 23\% of greenhouse gas emissions in the UK, with energy usage for heating responsible for the majority of emissions \cite{ccc2020SixthCarbonBudget}.
The design and management of building energy systems must be improved to reduce building energy usage and the incurred emissions in order to meet the UK's net-zero carbon emissions targets \cite{ccc2020SixthCarbonBudget}.
Building energy systems are subject to substantial uncertainties, such as future weather conditions which influence heating and cooling energy demands, occupant behaviour which affects the timing of energy usage, and building envelope characteristics which determine thermal response \cite{tian2018ReviewUncertaintyAnalysis}.
These uncertainties must be accounted for during design and operation in order to achieve energy efficiency \cite{yue2018ReviewApproachesUncertainty}.
Evidence, in the form of relevant data, can support decision making in the design and operation of building energy systems by reducing uncertainties, allowing more informed and effective decisions to be taken \cite{molina-solana2017DataScienceBuilding}.
However, data collection is costly. Many factors contribute to the overall cost of collecting and using data, such as sensor hardware, digital infrastructure to gather and store data, and computation required to process the data.
Therefore, the cost of data collection must be weighed against the benefits it provides to decision making to determine whether it is economically worthwhile.

The consideration of uncertainties during planning and the use of data collection to support decision making has been mainstream within the building energy literature for some time \cite{tian2018ReviewUncertaintyAnalysis,rysanek2013OptimumBuildingEnergy,mavromatidis2018ReviewUncertaintyCharacterisation,mobaraki2022NovelDataAcquisition}.
At the same time, data curation and data quality are widely acknowledged as significant issues due to high cost, maintenance, and sometimes poor reliability \cite{mobaraki2022NovelDataAcquisition,mantha2015RealTimeBuildingEnergy,xia2014ComparisonBuildingEnergy}. For example, \cite{han2020EnergysavingBuildingSystem} quotes the cost of sensors for occupancy detection to be \$0.2-3 per m$^2$ of building area, with many technologies suffering from either coverage or detection issues.
Substantial research attention has been devoted to reducing the quantity of data collection required in building energy systems, for example through the creation of standard datasets \cite{luo2022ThreeyearDatasetSupporting}, and to reducing the cost of data collection \cite{streltsov2020EstimatingResidentialBuilding}.
However, few studies that propose the use of collected data discuss its cost. Studies that do quantify data collection costs frequently report only the sensor cost \cite{han2020EnergysavingBuildingSystem,johnzhai2020AssessingImplicationsSubmetering}, omitting several components of the overall cost of data collection, such as the costs of digital infrastructure, data processing, maintenance, and quality assurance.

The benefits of data collection vary across building energy systems \cite{cho2019EnergyPerformanceAssessment} and the decisions the data is used to support. Hence, the value add of data collection options must be quantified for each system and compared to their costs to determine optimal data collection strategies, and assess whether additional data collection is worthwhile.
At present, questions of the economic merit of data collection are not commonly asked in the building energy literature, with very few studies quantifying the value of additional data collection.
Hence, there may exist significant unidentified wastage in building systems in the form of low insight monitoring. This wastage will likely grow if data collection strategies are not rationalised as building digitisation continues to expand.


\subsection{Assessment of data collection benefit for building energy systems}



A literature review identified only three existing studies which attempt to quantify the benefit of data collection in building energy systems. \cite{johnzhai2020AssessingImplicationsSubmetering} reviews case studies from the literature in which submetered building energy usage data was used to identify energy savings measures, and investigates the relationship between the depth of submetering, the resolution of the building energy data used for analysis, and the energy savings achieved. It finds that submetering energy usage at higher spatial resolution enables greater energy savings down to the equipment level, but that submetering within plant equipment (sensor level) did not lead to additional energy savings. Further, the average cost savings achieved at each submetering resolution are quantified, and compared to the average hardware cost of the metering system. However, whilst this study provides evidence to suggest that energy submetering is likely to be economically beneficial for the types of building energy systems covered in the reviewed case studies, it does not provide a methodology for determining whether a proposed metering system for a given building will be worth its cost. Further, as this approach relies on case studies, the conclusions cannot be used to assess the benefits of data collection for new systems, new data variables, or even existing systems under altered conditions.

The second study, \cite{winschermann2023AssessingValueInformation}, uses a simulation environment to quantify the impact of information availability on the performance of EV charging strategies for office buildings with respect to satisfying user charging demands and limiting power draw from the grid. It provides a relatively complete study, quantifying the improvements in charging performance achieved when historic data on EV charging behaviours for the specific office building are available for planning, including the consideration of combinations of data variables, and the case where perfect foresight of EV behaviour is used. However, the costs of acquiring and using this data to optimize EV charging are not quantified, nor is the charging performance of the energy system linked to a practical cost, e.g. the cost of grid connection or the lost revenue from unfulfilled charging demand. Therefore, the results cannot be used to guide charging system designers as to whether the use of historic data for planning is worthwhile. Additionally, whilst the simulation based approach is quite general, no clear framework for repeating the analysis to study different data collection strategies or problem contexts is provided.

A recent study, \cite{langtry2024ImpactDataForecasting}, investigates the impact of data usage on the prediction accuracy of forecasting models and the resulting operational performance of model predictive control achieved for scheduling battery storage in a simulated multi-building energy system. The benefit of collecting additional building metering data to support the control of the distributed energy system, and reduce the cost of meeting energy demands, was quantified indirectly via its impact on forecasting accuracy. The study finds that collecting more than 2 years of historic data does not provide a significant benefit for forecast accuracy and so battery scheduling, and that using metering data from existing buildings allows model predictive controls to be deployed without the collection of any data from the target building with only 10\% worse forecast accuracy. However, the reductions in operational cost enabled by data collection are quantified in terms of a relative objective function, meaning the economic benefit of additional data is not quantified, neither is the cost of obtaining that data. Though, guidance on how a system designer could conduct such an analysis is provided.


Questions of the benefits of data collection and how to determine data collection requirements have been explored in the context of building energy modeling and the calibration of building energy models. Studies have quantified the impacts of data usage on the accuracy of energy model calibration \cite{glasgo2017AssessingValueInformation,risch2021InfluenceDataAcquisition}, the relative sensitivity of models to energy data to prioritise data collection for calibration \cite{tian2016IdentifyingInformativeEnergy}, data precision requirements and the cost of data collection \cite{wang2022DataAcquisitionUrban}, and methods for maximizing the efficacy of collected data to reduce data collection costs \cite{han2021ApproachDataAcquisition}. However, the economic benefits of data collection for building energy model calibration cannot yet be quantified to determine whether its costs are worthwhile.

\subsection{Value of information analysis}



Value of Information analysis (VoI) \cite{raiffa1969ReviewDecisionAnalysis,howard1966InformationValueTheory} is a mature methodological framework from the Bayesian Decision Analysis field for quantifying the expected (mean) benefit of information acquisition to support decision making by reducing epistemic uncertainty \cite{zhang2021ValueInformationAnalysis}, which is the uncertainty associated with lack of knowledge of a system. It provides a general methodology for determining the benefit of data collection for any given decision problem, provided a mathematical description of the decision problem can be produced, i.e. the decision can be formulated as a stochastic decision problem \cite{pratt1995IntroductionStatisticalDecision}.
VoI has been widely used in fields such as medicine, agriculture, and environmental science \cite{keisler2014ValueInformationAnalysis} to quantify the economic value of information acquisition and determine optimal data collection strategies to support decision making. Within engineering, VoI has been used to study data collection strategies to support decisions regarding maintenance scheduling for buildings \cite{grussing2018OptimizedBuildingComponent} and wind farms \cite{myklebust2020ValueInformationAnalysis}, construction project planning \cite{esnaasharyesfahani2020PrioritizingPreprojectPlanning}, sensor placement \cite{malings2016ValueInformationSpatially}, and structural health monitoring \cite{difrancesco2021DecisiontheoreticInspectionPlanning,difrancesco2023SystemEffectsIdentifying}.


Only a single previous study applies VoI to decision problems within the building energy systems field. For a building energy system with distributed generation and storage, this study \cite{niu2023FrameworkQuantifyingValue} quantifies the expected reduction in the total lifetime cost (capital investment plus operational cost) achieved when the system is designed with perfect knowledge of the building energy demand and solar generation experienced during operation, compared to the case where the system is designed with uncertain knowledge of those operating conditions.
Whilst the study does not acknowledge the mathematical VoI framework or literature, it computes the Expected Value of Perfect Information (defined in Section \ref{sec:voi}) for the decision problem of designing a building energy system.
However, this quantification of the value of information is not linked to any practical data collection strategy. In fact, the uncertainties in operational conditions considered in the analysis are inherently not measurable as they pertain to a system that is yet to be designed or built. Therefore, the results obtained cannot provide any guidance to building energy system designers on the benefits of data collection to support system design, and whether expenditure on data collection is worthwhile.

The capabilities of VoI to provide a methodology for rationalising and prioritising expenditure on data collection to support decision making, and to determine optimal data collection strategies, have been demonstrated through its successful applications in a broad range of engineering fields. This motivates the use of VoI to study decision making in building energy systems, and address the unresolved questions of the economic merit of data collection in building energy systems, e.g. from building monitoring systems, raised in Section \ref{sec:background}.

\subsection{Research objectives \& novel contributions}

In the existing literature, few studies have sought to quantify the economic benefit of data collection for supporting decision making in building energy systems. In fact, no studies investigate whether data collection provides net economic benefit to decision making, i.e. whether the data is worth its cost.

\begin{enumerate} 
    \item[] This study proposes that Value of Information analysis is a methodology well suited to addressing this research gap. The main objective of the study is to demonstrate that the VoI framework can be applied to quantify the benefits of data collection in building energy systems, and so provide evidence for justifying, rationalising, and prioritising expenditure on data collection. This is achieved by illustrating the insights that VoI can provide for the following example decision problems:
\begin{itemize}
    \item Smart meters to support maintenance scheduling for air-source heat pumps
    \item Occupancy monitoring systems to support ventilation scheduling for indoor air quality in offices
    \item Ground thermal testing to support the design of ground-source heat pump based heating systems for apartment blocks
\end{itemize}
    \item[] {\color{\hlcolor}Each example problem demonstrates a different decision making context: management, operation, and design respectively; and illustrates a different aspect of VoI: representation of complex systems, sensitivity to model assumptions, and determination of optimal measurement precision.}
\end{enumerate}

This work is the first to apply Value of Information analysis to quantify the economic benefit of data collection to support decision problems in the context of building energy systems. Further, it provides the first numerical demonstration of the cost effectiveness of data collection, specifically the use of monitoring systems and field tests, to support the management, operation, and design of building energy systems.

{\color{\hlcolor}With a framework for quantifying the economic benefit of data collection options, building designers and managers will be able to economise and prioritise their expenditure on data. This will prevent wasteful collection of low insight data, and enable focus of resources on observations that improve decision making within building energy systems, reducing the overall cost of energy provision.}

The remainder of this work is structured as follows. Section \ref{sec:voi} briefly outlines the mathematical formulation of the VoI framework and the analyses that can be performed. Sections \ref{sec:ashp} through \ref{sec:gshp} present the three example decision problems. In each Section, the model of the decision problem is described, VoI is performed to study a different aspect of the benefits of data collection, and the insights provided to the decision maker are discussed. Finally, conclusions are drawn in Section \ref{sec:conclusions}.
\newpage
\section{\color{\hlcolor}Methodology} \label{sec:voi}

Value of Information analysis (VoI), as proposed by Raiffa \cite{raiffa1969ReviewDecisionAnalysis} and Howard \cite{howard1966InformationValueTheory} in the 1960s, is a framework based on Bayesian Decision Analysis and Expected Utility Theory \cite{smith1945TheoryGamesEconomic} for quantifying the expected benefit of information acquisition for making a given decision under uncertainty, due to the reduction in epistemic uncertainty (uncertainty associated with lack of knowledge of a system) provided by the information. This section provides a brief formulation and discussion of the VoI framework, to illustrate how decision making within building energy systems should be described to allow the benefits of data collection to be quantified using this methodology. A thorough explanation of the VoI framework is available in \cite{zhang2021ValueInformationAnalysis}.

\subsection{Bayesian Decision Analysis}

Bayesian Decision Analysis provides a mathematical framework for studying decision-making in the presence of uncertainties{\color{\hlcolor}, referred to as stochastic decision problems. It} seeks to determine the expected optimal action which should be taken by a decision-maker {\color{\hlcolor}(termed an `actor')} in order to maximise their expected utility, {\color{\hlcolor}that is, the decision which when taken in the system provides the highest reward/benefit to the decision-maker on average over the uncertainties in the problem. This task can be formulated as a mathematical (stochastic) optimization problem.}

Consider a generalised stochastic decision problem in which an actor seeks to select a `decision action' to take, $a \in \mathcal{A}$, within a system with uncertain parameters $\theta$, which have a prior probabilistic model {\color{\hlcolor}(distribution),} $\pi(\theta)$. The performance{\color{\hlcolor}/benefit} of each available action is given by a utility function which is also dependent upon the uncertain parameters, $u(a,\theta)$. In VoI analysis, before an action $a$ is taken, the actor may choose to take a `measurement action', $e \in E$, from which the actor receives data $z$. The probabilistic model describing the measurement data $f_e(z|\theta)$ is used to update the prior model, $\pi(\theta)$, to produce a posterior probabilistic model {\color{\hlcolor}(distribution)}, $\pi(\theta|z)$, which is then used by the actor to inform their choice of `decision action', and improve their decision making performance.

The set of available actions, prior probabilistic model, and utility function, $\lbrace\mathcal{A},\pi(\theta),u(a,\theta)\rbrace$, provide a complete mathematical description of the decision making task under uncertainty. The likelihood function $f_e(z|\theta)$ describes the reduction in epistemic uncertainty in the system, i.e. uncertainty in the parameters $\theta$, provided by data collection.

\begin{figure}[h]
    \centering
    \includegraphics[width=0.75\linewidth]{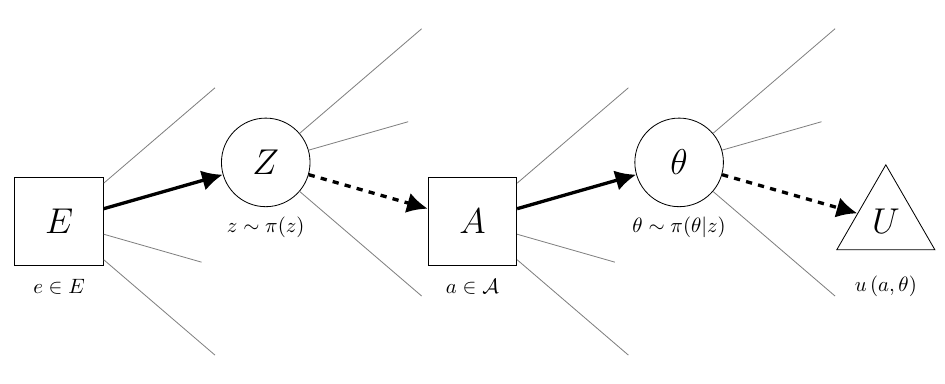}
    \caption{Decision tree representation of Pre-Posterior Decision Problem}
    \label{fig:DT-prepost}
\end{figure}

This generalised model can be represented {\color{\hlcolor}graphically} in decision tree form, as shown in Fig. \ref{fig:DT-prepost}, in which square nodes represent decisions, circular nodes represent uncertainties, and triangular nodes represent utilities.

The actor, who is assumed to be risk neutral, seeks to maximise their expected utility obtained from the selected `decision action'. Costs are defined as negative utilities. The actor may choose to do this without taking any measurement. The resulting stochastic optimisation is termed the Prior Decision Problem,
\begin{equation}
    \max_{a \in \mathcal{A}} \, \mathbb{E}_{\theta} \left\lbrace u(a,\theta) \right\rbrace
\end{equation}

Alternatively, the decision-maker may additionally consider the selection of a `measurement action' within the decision variables of the optimisation, which leads to the Pre-Posterior Decision Problem,
\begin{equation}
    \max_{e \in E} \, \mathbb{E}_{z} \left\lbrace \max_{a \in \mathcal{A}} \, \mathbb{E}_{\theta|z} \left\lbrace u(a,\theta) \right\rbrace \right\rbrace
\end{equation}


\subsection{Value of Information}

The expected Value of Information (VoI) \citep{raiffa1969ReviewDecisionAnalysis} can be computed using the Bayesian Decision Analysis framework, and is defined as the increase in the expected utility achieved by the optimal action taken by the actor as a result of having additional information at the time of selecting the `decision action'.

In the general case where a measurement $e$ provides imperfect/uncertain information that reduces the epistemic uncertainty in the system, the expected value of that uncertain information is termed the Expected Value of Imperfect Information (EVII), and is given by,

\begin{align} \label{eq:EVII}
\begin{split}
    \begin{split}
    \text{EVII}(e) &= \mathbb{E}\lbrace \text{imperfect information decision utility} \rbrace \\
        &\qquad\qquad - \mathbb{E}\lbrace \text{prior decision utility} \rbrace
    \end{split} \\[0.5em]
        &= \mathbb{E}_{\theta} \left\lbrace \max_{a \in \mathcal{A}} \: \mathbb{E}_{\theta|z} \left\lbrace u(a,\theta) \right\rbrace \right\rbrace - \max_{a \in \mathcal{A}} \: \mathbb{E}_{\theta} \left\lbrace u(a,\theta) \right\rbrace \\[1em]
        &= \mathbb{E}_{\theta} \left\lbrace \max_{a \in \mathcal{A}} \: \mathbb{E}_{\theta|z} \left\lbrace u(a,\theta) \right\rbrace - u(a^*,\theta) \right\rbrace
    \end{split}
\end{align}

where the posterior probabilistic model, $\pi(\theta|z)$, used to compute inner expectation with respect to $\theta|z$, is derived using the likelihood function for the measurement $e$, $f_e(z|\theta)$.

In the case where measurement provides perfect information on the true value of the uncertain parameters of the system, i.e. $\pi(\theta|z) = \delta(\theta{-}z)$, the expected utility obtained by the actor making decisions with perfect information simplifies as,
\begin{equation}
\mathbb{E}_{z} \left\lbrace \max_{a \in \mathcal{A}} \, \mathbb{E}_{\theta|z} \left\lbrace u(a,\theta) \right\rbrace \right\rbrace \: \longrightarrow \: \mathbb{E}_{z} \left\lbrace \max_{a \in \mathcal{A}} \, u(a,z) \right\rbrace
\end{equation}
The expected value of information in this case is termed the Expected Value of Perfect Information (EVPI), and is given by,
\begin{equation} \label{eq:EVPI}
    \begin{aligned}
        \text{EVPI} &= \mathbb{E}\lbrace \text{perfect information decision utility} \rbrace \\
        & \qquad\quad - \mathbb{E}\lbrace \text{prior decision utility} \rbrace \\
        &= \mathbb{E}_{z} \left\lbrace \max_{a \in \mathcal{A}} \, u(a,z) \right\rbrace - \max_{a \in \mathcal{A}} \mathbb{E}_{\theta} \left\lbrace u(a,\theta) \right\rbrace \\[1em]
    \end{aligned}
\end{equation}
The EVPI is significantly cheaper to compute\footnote{The EVPI is quadratically cheaper to compute compared to the EVII in the number of evaluations of the utility function used to compute the expectations. This is highly advantageous when the utility function computes computationally expensive to evaluate, e.g. where it requires simulation of an engineering system.}, and provides an upper-bound on the EVII \cite{keisler2014ValueInformationAnalysis}. 

The Value of Information can be interpreted as the increase in the expected utility the actor can achieve as a result of having improved knowledge of the uncertain parameters of the system, i.e. reduced epistemic uncertainty. {\color{\hlcolor}That is, on average how much better the decision-maker is able to do at making their decision as a result of having additional information provided by a measurement.} Alternatively it can be viewed as the reduction in expected utility arising from the presence of the additional uncertainty in the system.
Typically, decision problems are formulated using an economic objective, i.e. a total cost or profit. Through the former interpretation, the VoI quantifies the actor's willingness to pay for information that reduces epistemic uncertainty in the parameters under study from the decision problem. Comparing VoI to the cost of obtaining information allows the net economic benefit to be quantified, i.e. the VoI minus the cost of information. With this, a decision maker can determine whether the information acquisition considered is economically worthwhile, and compare the relative benefit of different data collection options.

For systems where the uncertain parameters are modelled using discrete probability distributions, the expectations required for VoI calculations can be evaluated exactly. However, when there are continuously distributed uncertain parameters in the system, as is common in practical engineering problems, these expectations must be estimated, e.g. using Monte Carlo methods or by discretizing the parameter space.

\subsection{Influence diagrams}

{\color{\hlcolor}While decision trees provide a complete graphical illustration of the possible outcomes of a stochastic decision problem, f}or decision problems comprising larger numbers of nodes with more complex dependency networks, or problems which contain parameters in continuous spaces, influence diagram{\color{\hlcolor}s are used instead to provided a simpler and clearer description of the structure of a decision problem} \citep{difrancesco2023GuidanceUseProbabilistic}. Influence diagrams depict only the causal dependencies between nodes, and not realisation trajectories through the decision graph. This allows experts to readily describe the structure of decision making in complex engineering systems, and provides a far clearer description of complex decision problems. In some instances whilst there may exist a decision tree compatible description of the problem, an alternative description containing more nodes and causal dependencies is chosen to provide insight into the physics or dynamics of the system, and so an influence diagram is used. This is the case for the decision problems studied in Section \ref{sec:experiments}.

\subsection{Analysis extensions}

The standard VoI framework assumes decision makers to be risk neutral, and does not quantify or account for risk during decision making, studying only the impact of data collection on the expected outcome of the decision problem. However, in many practical engineering systems, risk has a significant influence on decision making, and data collection can provide value in its ability to reduce the risk of decisions. Risk aversion has been studied in the context of building energy system design through the inclusion of risk measures in the design objective, such as the Conditional Value at Risk \cite{pickering2021QuantifyingResilienceEnergy}. Whilst the standard VoI framework models the decision utility as dependent only on the action taken and true underlying value of the uncertain parameters, it can be readily extended to allow utility functions that are dependent on the distribution of outcomes and account for the cost of risk. Additionally, measures of the risk associated with the decisions taken can be computed within the standard framework, as the distributions of outcomes are available during the VoI calculation procedure. 

A common critique of the VoI framework is that the VoI values computed are valid only under the particular formulation of decision problem studied. The VoI results do not generalise and can change substantially under perturbations of the problem setup. Whilst this issue of generalisation exists in all complex modeling tasks, VoI outputs can be particularly sensitive to models assumptions. Seemingly innocuous changes in probability distributions can result in dominant actions which cause the VoI to fall to zero.
Sensitivity analysis can be performed to validate the results obtained with respect to the assumptions made in the formulation of {\color{\hlcolor}models of the system and uncertainties}, and address the issue of generalisation by determining the space of problem setups for which the results hold. However, sensitivity analyses are computationally expensive as the VoI analysis must be repeated for each problem setup considered.

The VoI framework can also be extended to quantify the improvement in expected utility derived from accounting for stochasticity during decision making compared to the case where planning is performed using a deterministic model of the system, termed the Value of Stochastic Solution \cite{birge1982ValueStochasticSolution}, and the study of the relative benefits of combinations of data collection activities, termed `system effects' \cite{difrancesco2023SystemEffectsIdentifying}.
\section{Example decision problems} \label{sec:experiments}

Three example decision making problems are presented, covering the management, operation, and design of building energy systems. Value of Information analysis is applied in each case, and the insights VoI provides into the benefits of data collection to support decision making are discussed. Each example illustrates a different aspect of VoI, including the representation of complex engineering systems, the impact of model assumptions on VoI, and the determination of optimal measurement precision for data collection. {\color{\hlcolor}These example problems examine simplified models of buildings, uncertainties, and decision making. Their aim is to demonstrate how the VoI methodology can be applied to building energy systems to provide insight into the role of data collection across a broad range of contexts.} All code used to perform the example VoI calculations, {\color{\hlcolor}and all numerical results used to create the figures in this article, are} available at \censor{\href{https://github.com/EECi/VOI-for-Building-Energy}{\nolinkurl{github.com/EECi/VOI-for-Building-Energy}}}.

\subsection{Condition Measurement for Optimising Maintenance Scheduling for Air-Source Heat Pumps} \label{sec:ashp}

Air-source heat pumps (ASHPs) provide significant advantages to decarbonising the heating of buildings through their ability to exploit ambient heat in the environment to achieve high Coefficients of Performance (COPs), reducing the direct energy input required to heat a space, and so reducing the embodied carbon emissions of heating. However, through usage, the performance of ASHPs degrades, reducing the COPs they can achieve. Maintenance activities can be undertaken to address this performance degradation and improve the COPs achieved by the heat pumps, reducing the electricity consumed and operational costs. However, regular maintenance is costly.

When deciding how frequently to maintain ASHP units, asset owners seek to trade-off the cost of maintenance activities with the benefits they provide in reduced electricity consumption cost to minimise the total cost of operating the ASHP asset. However, the performance of degrading ASHPs depends on several uncertain factors, including the degradation rate, the base Seasonal Performance Factor (SPF), and the performance improvement provided by maintenance. Additionally, at the time maintenance is scheduled, the cost of electricity and the heating load of the building are unknown. Therefore, the maintenance scheduling decision must be made under these uncertainties.

Smart meter data can be used to better estimate the performance of ASHP units, allowing for the scheduling of maintenance to be optimised. However, installing and maintaining smart meters adds additional cost to the operation of the ASHP units. Therefore, asset owners will raise the question, ``Does installing a smart meter on an ASHP unit reduce the overall operating costs by allowing for optimised maintenance scheduling?''.\\

A case involving a research building at \censor{the University of Cambridge}, which is equipped with 30 identical ASHP units, is examined to assess the economic viability of {\color{\hlcolor}installing a centralised electricity smart meter to assist} maintenance scheduling optimisation. The asset owner selects the number of evenly spaced maintenance activities undertaken per year, $N_m \in \{0,\ldots,12\}$, as to minimise the expected {\color{\hlcolor}operating cost of the heating system (Eq. \ref{eq:ASHP-total-cost})}.\\

The annual energy consumed by the ASHP units is given by,
\begin{equation}
    E = \frac{L_H}{\text{SPF}}
\end{equation}
where $L_H$ is the heating load of the building for the coming year. This heating load is assumed to be Gaussian distributed, with mean and standard deviation determined from historic building metering data,
\begin{equation}
    L_H \sim \mathcal{N}\left( L_H, \mu=12.6, \sigma=1.36 \right) \:\: \text{GWh/year}
\end{equation}

The total cost of electricity consumed to meet the building heating load is given by,
\begin{equation}
    C_e = p_e E
\end{equation}
where $p_e$ is the price of electricity for the coming year, which is assumed Gaussian with mean and standard deviation calculated using consumer electricity cost data across UK regions from 2022 \cite{departmentforenergysecurityandnetzero2023AnnualDomesticEnergy},
\begin{equation}
    p_e \sim \mathcal{N}\left( p_e, \mu=32.6, \sigma=1.6 \right) \:\: \text{p/kWh}
\end{equation}

The annual Seasonal Performance Factor ($\text{SPF}$) of the ASHPs, the average COP over the heating season, is given by,
\begin{equation}
    \text{SPF} = \text{SPF}'(1-\alpha)(1+\beta)
\end{equation}
where $\text{SPF}'$ is the base heat pump SPF, which is not known at the time of installation. Using data from existing ASHP units \cite{nouvel2015EuropeanMappingSeasonal}, it is assumed to be distributed as,
\begin{equation}
    \text{SPF}' \sim \mathcal{N}\left( \text{SPF}', \mu=2.9, \sigma=0.167 \right)
\end{equation}
$\alpha$ is the performance degradation factor of the ASHPs, which is modelled as being distributed as a truncated Normal with mean 0.01 and standard deviation 0.25, as it is a non-negative parameter,
\begin{equation}
    \alpha \sim \mathcal{N}(\alpha, \mu=1e{-}2,\sigma=0.25 : \alpha \geq 0)
\end{equation}
$\beta$ models the performance improvement provided by maintenance activities, and is given by,
\begin{equation}
    \beta = \frac{\beta_a N_m^\gamma}{\beta_b + N_m^\gamma} (1+\varepsilon)
\end{equation}
the parameters $\beta_a$, $\beta_b$, $\gamma$, are empirical parameters with assumed values of 0.05, 2.5, and 1.4 respectively \cite{griffith2008MethodologyModelingBuilding}. The parameter $\varepsilon$ models uncertainty in the performance improvement, and is taken to be distributed as,
\begin{equation}
    \varepsilon \sim \mathcal{N}(\varepsilon, \mu=0,\sigma=0.1)
\end{equation}
The cost {\color{\hlcolor}per activity of performing maintenance on all 30 ASHP units}, $C_m$, is taken to be £18,000 \citep{daikin2022DaikinUKPrice}. So, the total {\color{\hlcolor}cost of operating the ASHPs (comprised of electricity and maintenance costs)} is given by,
\begin{equation} \label{eq:ASHP-total-cost}
    C_{\text{total}} = C_e + C_m N_m
\end{equation}

\begin{figure}[h]
    \centering
    \includegraphics[width=0.6\linewidth]{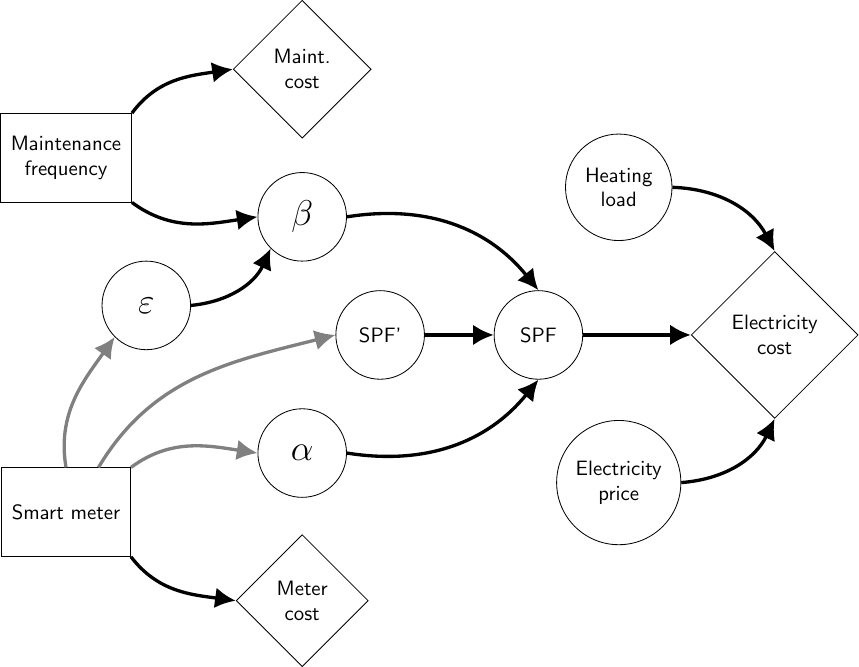}
    \vspace{2pt}
    \caption{Influence diagram representation of ASHP maintenance scheduling decision problem}
    \label{fig:ID-ASHP-maintenance}
\end{figure}

The described stochastic decision problem of optimally scheduling ASHP maintenance is represented in influence diagram form in Fig. \ref{fig:ID-ASHP-maintenance}, where grey arrows represent measurements. This diagrammatic representation provides a far clearer, more concise, and readily extendable description of the decision task within this complex engineering system.

\begin{figure}[!h]
    \centering
    \begin{minipage}{.475\textwidth}
        \centering
        \includegraphics[width=\linewidth]{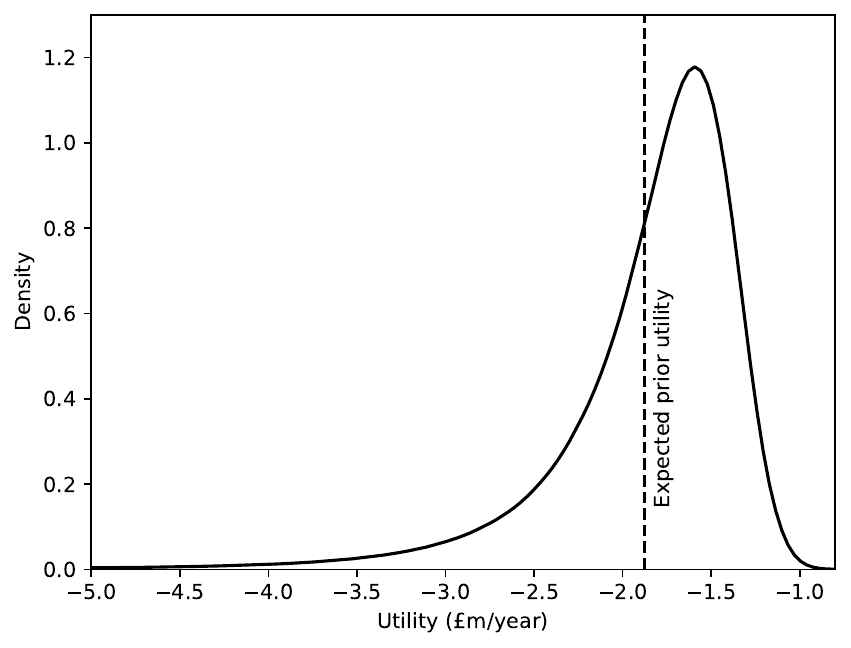}
        \caption{Distribution of utilities achieved by optimal prior action, ${N_m}^*=2$. Dashed line indicates mean of distribution}
        \label{fig:ASHP-prior-dist}
    \end{minipage}%
    \hfill
    \begin{minipage}{.475\textwidth}
        \centering
        \includegraphics[width=\linewidth]{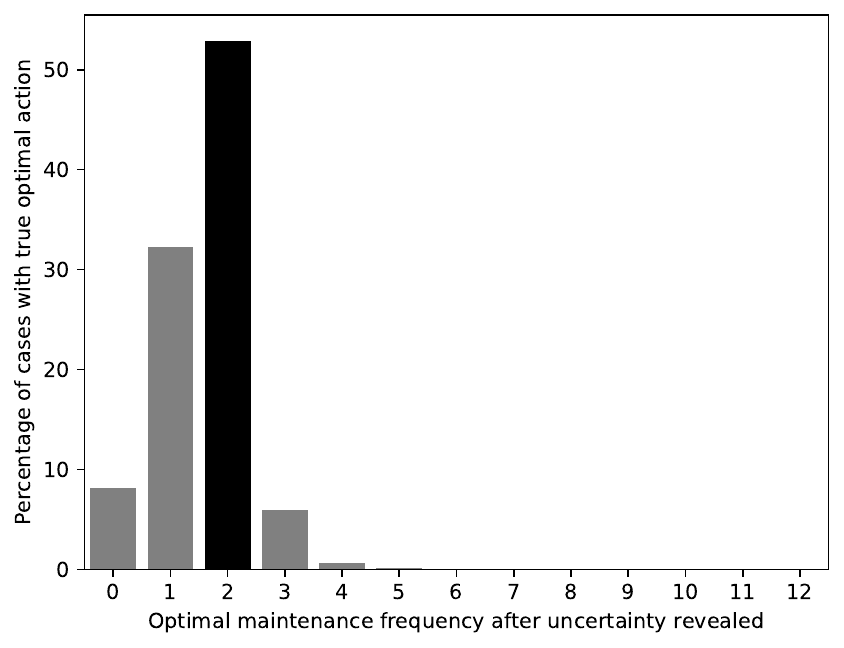}
        \caption{Distribution of true optimal maintenance frequency, i.e. best action once uncertainty has been revealed}
    \label{fig:ASHP-post-action-freqs}
    \end{minipage}%
\end{figure}

Solving the Prior Decision Problem, the optimal maintenance frequency is found to be 2 activities per year, leading to an expected {\color{\hlcolor}operating cost (Eq. \ref{eq:ASHP-total-cost})} of £1,876,300/year for the building heating system{\color{\hlcolor}, of which £36,000/year (2\%) is spent on maintenance}. Fig. \ref{fig:ASHP-prior-dist} plots the probability distribution of operational costs for the system when maintained 2 times a year. It demonstrates the significance of the uncertainties in the building energy system on the planning problem, as the cost of operating the heating system can vary from £1m/year to £4m/year, and so highlights the importance of accounting for those uncertainties during planning.

Solving the Pre-Posterior Decision Problem with perfect information, i.e. planning maintenance with perfect knowledge of the uncertain parameters in the system, achieves an expected cost of £1,874,640/year. Therefore the Expected Value of Perfect Information (EVPI), as defined in Eq. \ref{eq:EVPI}, is £1,660/year.

Fig. \ref{fig:ASHP-post-action-freqs} shows the distribution of optimal maintenance frequency when the true values of the uncertain system parameters are known. In slightly over half of cases, perfect information about the system does not change the maintenance scheduling decision, i.e. the true solution is the same as the prior solution, indicated by the black column. However the information provides value, as in many cases the performance degradation of the ASHPs is less severe, with less frequent maintenance providing lower operational cost. Monitoring information allows the system operator to avoid the cost of unnecessary maintenance activities in these cases.

The {\color{\hlcolor}annualised cost of the electricity smart meter used to monitor} the ASHP units {\color{\hlcolor}(`meter cost' in Fig. \ref{fig:ID-ASHP-maintenance})} is estimated to be £{\color{\hlcolor}70}/year \citep{daikin2022DaikinUKPrice}. Therefore, the installation of smart meters capable of perfectly measuring ASHP performance, degradation, and building load, and the dynamic scheduling of maintenance activities using the actual electricity price, would lead to a net economic benefit, VoI minus information cost, of £{\color{\hlcolor}1,590}/year to the asset owner from reduced operational costs. In this way, the asset owner can justify their investment in smart meters.
The value of information provided by smart meters in this scenario is small compared to the total operational costs, only 0.06\%. However, the use of smart meters to optimise maintenance scheduling can provide additional benefits, such as extending the lifespan of the ASHP units and decreasing the probability of malfunctions. These cost savings, although not accounted for in these calculations due to simplification, could be significant.

\newpage

\subsection{Building Occupancy Measurement for Real-Time Ventilation Scheduling to Improve Indoor Air Quality} \label{sec:vent}

In mechanically ventilated office spaces, building managers must schedule ventilation system settings to ensure sufficient indoor air quality for the occupants. Adequate ventilation is required to prevent the transmission of airborne infectious diseases such as the SARS-CoV-2 virus (COVID-19), which is both damaging to the health of the occupants and costly for the tenant of the office space through lost productivity. However, operating ventilation at an unnecessarily high rate can impact occupant thermal comfort, lead to excessive carbon emissions, as well as excessive operational cost of the ventilation system through the additional heating/cooling demand required to condition air intake. The risk of viral transmission, and thus the appropriate ventilation setting, is highly dependent on the number of occupants in the office space. However, in the absence of occupant monitoring and dynamic ventilation control systems, building managers must schedule ventilation system settings without knowledge of the exact occupancy level of the space. At the time of deciding ventilation scheduling, the occupancy of the ventilated space over the operation period is uncertain. This uncertainty is particularly relevant in light of recent trends towards `work from home', which reduces the predictability of office occupancy.

This raises the question, ``Would it be worth installing a smart occupancy monitoring and ventilation control system in an office space to improve ventilation scheduling?'', i.e. would the economic benefit of improved ventilation control be greater than the cost of installing such a smart monitoring and control system?\\

A simple model of indoor air quality in a typical office space is considered. Said office space is taken to have a maximum occupancy of 100 people, a floor area of 1,000 m$^2$, and a ceiling height of 2.4 m. There are five available ventilation settings: 1, 3, 6, 12, and 20 air changes per hour (ACH). The ventilation system is assumed to have a fan of specific power 1.9 W/l/s, $P_{\text{fan}}$, operating at 60\% efficiency, $\eta$, for 10 hours per day, $T$. At an electricity unit price of 32.6 p/kWh, $p_e$, the cost of operating the ventilation system\footnote{The cost of heating/cooling associated with ventilation is neglected for simplicity, but can be readily added to the model.} at each setting is calculated as,
\begin{equation}
    C_{\text{vent}} = \text{ACH} \times V \times \frac{P_{\text{fan}}}{\eta} \times T \times p_e
\end{equation}
where $V$ is the volume of the office in litres, $T$ is expressed in seconds and $p_e$ in £/Wh.

A model of the probability of viral infection for an individual in an indoor space as described in \citep{deoliveira2021EvolutionSprayAerosol} is used, which is also available in web application form at \href{https://airborne.cam/}{\nolinkurl{airborne.cam}} \citep{gkantonas2021AirborneCamRisk}. It is assumed that the base prevalence of infection amongst the occupants is that of the general UK population in February of 2023, 2.18\% \citep{ons2023CoronavirusCOVID19Latest}, and that infection of an individual leads to 3 days of sick leave, which taking the median daily salary for full-time employees in 2022, costs the tenant £128/day \citep{ons2022EmployeeEarningsUK} in lost productivity.
It is further assumed that any occupants are present in the office space for the whole 8 hour work day, that time coupling effects between days can be neglected, i.e. that the model for a single day is representative, and so illness costs are calculated using the expected number of infections for the given occupancy. The prior distribution of occupancy is taken to be discrete uniform in the interval 0 to 100 inclusive.

The stochastic decision problem is thus to select the ventilation system setting which minimises the sum of the ventilation system operation (electricity) cost, and the cost of lost productivity due to illness to the tenant, subject to uncertainty in the occupancy level of the space.
\begin{equation}
    C_{\text{total}} = C_{\text{vent}}(\text{ACH}) + n \times p_{\text{infection}}(n,\text{ACH})
\end{equation}
where $n$ is the number of occupants in the office.

\begin{figure}[p]
    \centering
    \includegraphics[width=0.6\linewidth]{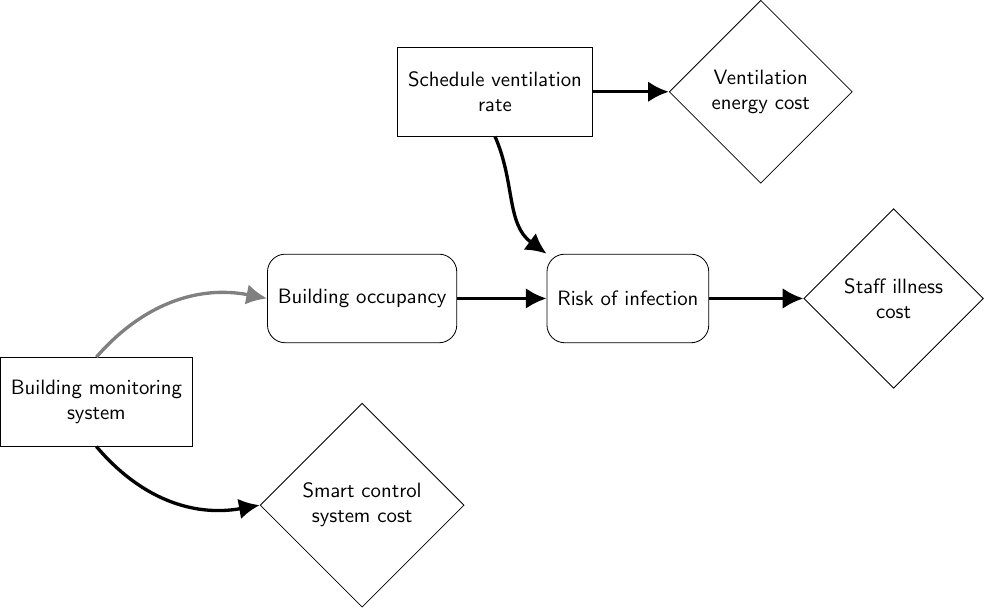}
    \vspace{4pt}
    \caption{Influence diagram representation of building ventilation scheduling decision problem} \label{fig:ID-building-vent}
\end{figure}

This decision problem can be described via the influence diagram provided in Fig. \ref{fig:ID-building-vent}.\\

\begin{figure}[p]
    \centering
    \begin{minipage}{.475\textwidth}
        \centering
        \includegraphics[width=0.975\linewidth]{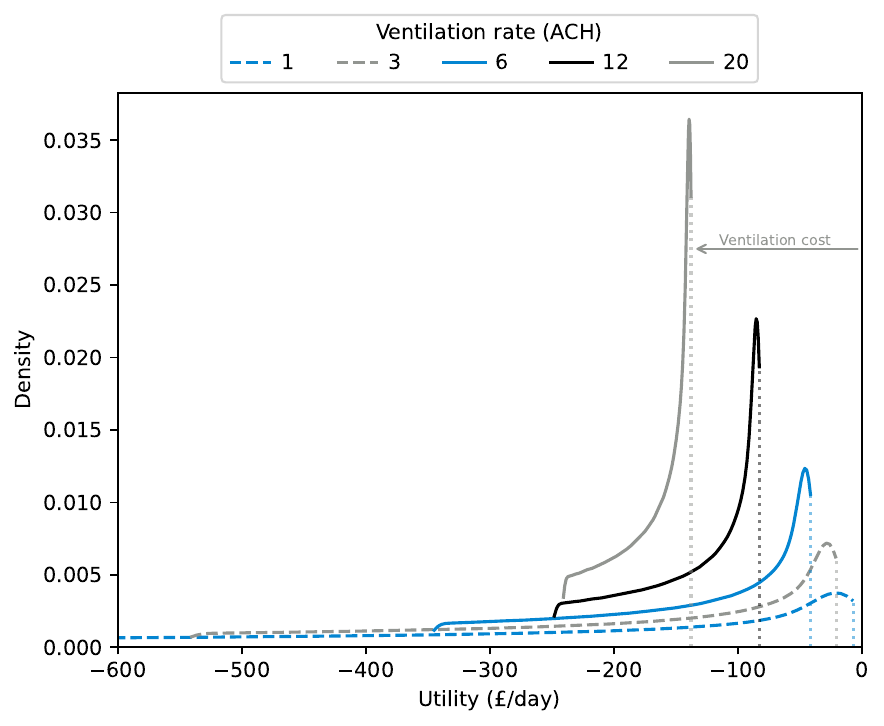}
        \caption{Distribution of utilities achieved by each ventilation rate under prior uncertainty}
        \label{fig:b_vent_a_dists}
    \end{minipage}%
    \hfill
    \begin{minipage}{.475\textwidth}
        \centering
        \includegraphics[width=0.95\linewidth]{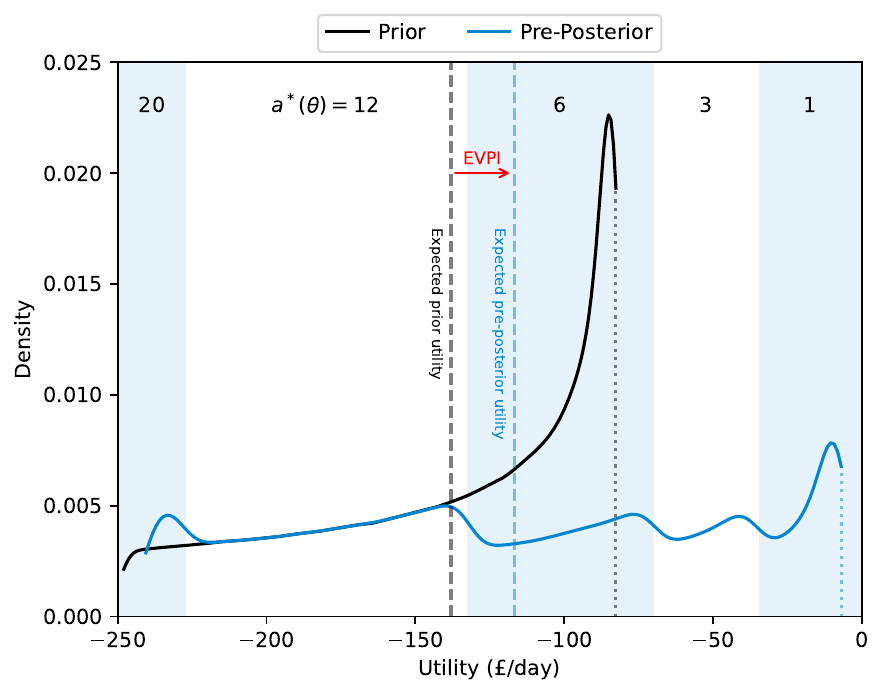}
        \caption{Distribution of utilities achieved by prior and pre-posterior decisions}
    \label{fig:b_vent_prior_vs_prepost}
    \end{minipage}%
\end{figure}

\begin{figure}[p]
    \centering
    \includegraphics[width=0.45\linewidth]{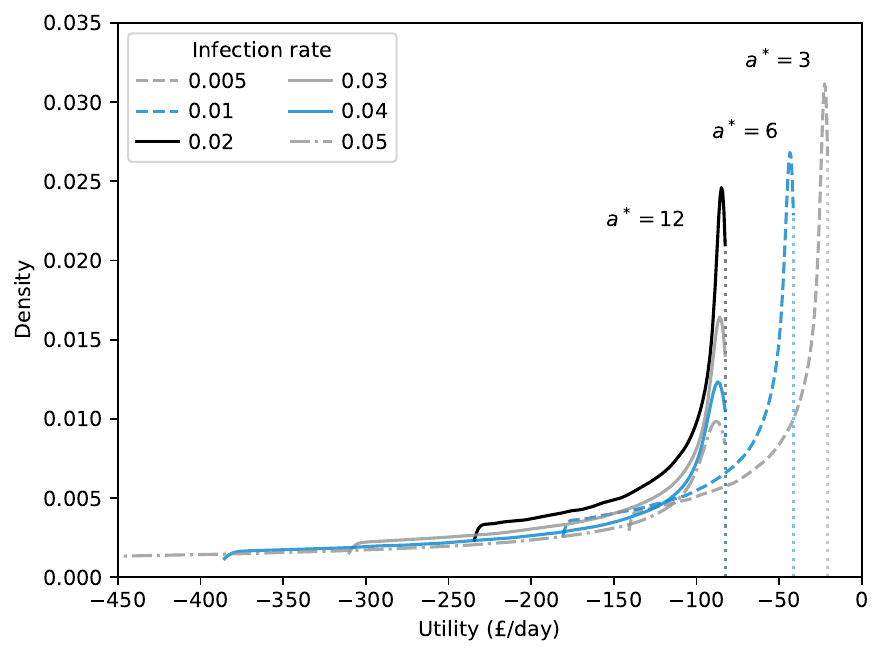}
    \caption{Distributions of utilities for prior solution with varying assumed infection rate} \label{fig:b_vent_sensitivity}
\end{figure}

Solving the Prior Decision Problem, the optimal ventilation rate is found to be 12 air changes per hour, leading to an expected overall cost of £138/day for the tenant. Fig. \ref{fig:b_vent_a_dists} shows the distribution of utilities for each available ventilation rate under the prior uncertainty. Increasing the ventilation rate increases the base cost of all outcomes, indicated by the dashed line showing the minimum cost for each curve, but reduces the risk of large numbers of infections occurring when the office has high occupancy, limiting the maximum cost. The utility distribution for the prior optimal ventilation rate of 12 ACH in the base case is indicated by the black curve in Figures \ref{fig:b_vent_a_dists} to \ref{fig:b_vent_sensitivity}.

Solving the Pre-Posterior Decision Problem with perfect information, i.e. dynamically scheduling ventilation with perfect knowledge of occupancy from a monitoring system, achieves an expected cost of £117/day. Therefore, the EVPI is £21/day. Hence, over a 20 year operational lifetime, the smart monitoring and control system could save the tenant up to £850,000, or 15\% of the total operating cost. Fig. \ref{fig:b_vent_prior_vs_prepost} plots the distribution of utilities achieved by the prior and pre-posterior decisions. For the prior, the same decision of 12 ACH is used in all cases as no measurement is taken. Whereas, for the pre-posterior, the ventilation rate is set using information on the occupancy level in each case. The blue regions indicate the regime of cases where each ventilation rate is selected by the pre-posterior decision, i.e. where it is the lowest cost choice. The pre-posterior density is similar to the prior in the region of high costs, but is much more uniform and extends substantially further into the low cost region, indicating that the VoI is primarily derived from the ability to reduce ventilation rate and save electricity cost when office occupancy is low.

Building managers can {\color{\hlcolor}use} the computed EVPI to support their decision making about whether investment in smart occupancy monitoring and ventilation control systems are worthwhile. If a proposed smart monitoring and control system is expected to have an installation and maintenance cost over the 20 year lifetime of more than £850,000, then this investment will not be beneficial for the context of ventilation scheduling. A cheaper but less precise method of estimating occupancy, such as a desk booking system, may be a more suitable strategy, or it may be most cost effective to forego occupancy measurement altogether and use a fixed ventilation rate.

\subsubsection*{Sensitivity analysis}
The problem formulation considered assumes a variety of properties about the office being ventilated. However, commercial building stocks are diverse, and the actual value of these properties varies between buildings. A sensitivity analysis is performed to investigate whether occupancy monitoring and dynamic ventilation scheduling remains valuable across different sized offices. The VoI analysis is repeated for office models with 5 to 25 m$^2$ of floor area per person, maintaining a maximum occupancy of 100, and keeping all other properties as before. Table \ref{tab:floor-area-sensitivity} provides the optimal ventilation rate without measurement and the EVPI for each size of office considered. The EVPI is lowest in the extreme cases where extreme ventilation rates are optimal both on average (for the prior problem) for the majority of occupancy outcomes. For instance, in the smallest office, the highest available ventilation rate of 20 ACH is the optimal prior decision, and the optimal decision for 53\% of occupancy outcomes. For all office sizes tested, the EVPI remains significant, 10\% or more of the prior expected cost, suggesting that occupancy monitoring systems are likely beneficial for offices of any size.

\begin{table}[h]
    \centering
    \renewcommand{\arraystretch}{0.75}
    \begin{tabular}{c|ccccc} \toprule \toprule
        Floor area per person (m$^2$) & \: 5 \: & \: 10 \: & \: 15 \: & \: 20 \: & \: 25 \: \\ \midrule
        Prior optimal ventilation rate (ACH) & 20 & 12 & 6 & 3 & 3 \\
        EVPI (£/day) & 14 & 21 & 16 & 19 & 14 \\
        \bottomrule \bottomrule
    \end{tabular}
    \caption{Sensitivity of prior solution and EVPI to office size}
    \label{tab:floor-area-sensitivity}
\end{table}

The prevalence of viral illness in the general population has a significant impact on the risk of infection spreading in an office environment, and so should be taken into account by a building manager when setting ventilation rates, and as a result when determining what information is required to support that decision. The sensitivity of the prior decision and EVPI is tested for infection rates from 0.5\% to 5\% \cite{ons2023CoronavirusCOVID19Latest}. Table \ref{tab:inf-rate-sensitivity} provides the results of this analysis, and Fig. \ref{fig:b_vent_sensitivity} plots the distribution of utilities for the prior solutions at each infection rate. As infection rate increases, the risk of infection at a given occupancy level rises, causing a widening of the utility distribution, and so greater risk for the decision maker. Additionally, the distribution of true optimal ventilation rates flattens. Hence, for more occupancy outcomes a different ventilation rate to the prior is optimal, and the benefit of that improved decision increases. Therefore, the greater the infection rate, the higher the VoI, and the more valuable occupancy monitoring becomes. {\color{\hlcolor}The sharp increase in EVPI between infection rates of 1\% and 2\% highlights the importance of using sensitivity analysis to determine whether conclusions about the economic benefits of data collection derived from VoI analysis remain valid as assumptions about the studied building energy system are varied. Alternatively, assumed parameters can be modelled as uncertainties and included within the system model.}

\begin{table}[h]
    \centering
    \renewcommand{\arraystretch}{0.75}
    \begin{tabular}{c|cccccc} \toprule \toprule
        Infection rate (\%) & 0.5 & 1 & 2 & 3 & 4 & 5 \\ \midrule
        Prior optimal ventilation rate (ACH) & 3 & 6 & 12 & 12 & 12 & 12 \\
        EVPI (£/day) & 8 & 11 & 22 & 21 & 24 & 30 \\
        \bottomrule \bottomrule
    \end{tabular}
    \caption{Sensitivity of prior solution and EVPI to infection prevalence in the population}
    \label{tab:inf-rate-sensitivity}
\end{table}

\newpage

\subsection{Ground Conductivity Measurement for Optimising Borehole Design in Residential Ground-Source Heat Pump Heat Supply Systems} \label{sec:gshp}

Ground-source heat pump (GSHP) systems use the ground as a source and sink of heat to provide cooling and heating for buildings in a highly energy efficient way \cite{aresti2018ReviewDesignAspects}. As these systems exchange heat with the ground, their performance (characterised by the COPs they achieve) is influenced significantly by the geological properties of the ground with which they exchange heat. In the design of such GSHP heating \& cooling supply systems, it is desired to match the capacity of the GSHP system to energy demand of the building as to minimise the overall cost of operating the supply system over its lifetime. The overall operating cost is composed of the capital cost of constructing the system, and the operational costs, which are primarily the cost of the electricity required to meet the building energy demands. Under-specification of the systems leads to greater electricity usage from less efficient and so higher cost auxiliary heating systems, whilst over-specification of the system results in unnecessary capital cost.

At the time of system design, the thermal properties of the ground are typically not known precisely, as existing geological survey data provides an uncertain estimate of the ground properties in the site location \cite{dallasanta2020UpdatedGroundThermal}. However, the system designer has the option to commission thermal tests at the site location prior to designing the GSHP system. Whilst these tests reduce the uncertainty in the ground thermal properties, they are time-consuming and incur additional costs. As a result, these tests are often not conducted, and generally available information on materials and location is used to provide uncertain estimates of ground thermal properties which are used to design the GSHP system. This therefore poses the following question to the designer, ``Would commissioning thermal tests reduce the overall lifetime cost of the GSHP heat supply system by improving the matching of the designed system capacity to the building load?''.\\

A simplified GSHP system design task for a residential building heat supply system is considered. In this design task, the designer must select the length of boreholes, $L_{\text{bh}}$, to be drilled for the ground heat exchange. The available length choices are 110 to 190 m, in 5 m increments. The capital cost of borehole drilling is taken to be 70 £/m/borehole.

It is assumed that the effective ground thermal conductivity, $\lambda_{\text{ground}}$, is the only uncertain geological parameter, and that it is Normally distributed with mean 1.94 W/mK, and standard deviation 0.31 W/mK,
\begin{equation}
    \lambda_{\text{ground}} \sim \mathcal{N}(\mu=1.94,\sigma=0.31) \:\: \text{W/mK}
\end{equation}
which is the average thermal conductivity over the range of borehole depths considered, and covers typical uncertainty ranges given the heterogeneity present in soils \cite{busby2018ModellingStudyVariation,busby2011ProvisionThermalProperties,loveridge2013ThermalResponseTesting}.

The designed GSHP system, consisting of 12 boreholes, must supply heat to a small apartment block of 10 flats over a 50 year operational lifetime. The building is assumed to have a typical heating demand distribution for its type, and a simplified load profile is synthesised based on demand values for the UK \citep{mitchell2020UKPassivhausEnergy}, and using historic weather data for London. This synthetic load profile, $E_{\text{load}}(t)$, consumes 116MWh/year, {\color{\hlcolor} corresponding to a 13.2 kW mean heating load for the building}, with a peak load power of 30.5 kW. The data can be found in the \href{https://github.com/EECi/VOI-for-Building-Energy}{GitHub repository}.

The model of geothermal borehole operation presented in \citep{lamarche2007NewContributionFinite} is used to determine the scheduling of energy extracted from the ground in each time instance, $E_g(t)$, with the borehole fluid temperature, $T_{\text{fluid}}$, conservatively constrained to be within the range $5$ degC to $35$ degC,
\begin{equation}
    5 \leq T_{\text{fluid}}(t) \leq 35
\end{equation}
In this model, $T_{\text{fluid}}$ is a function of the power extracted from the ground, $E_g(t)$, the ground thermal conductivity, $\lambda_{\text{ground}}$, the borehole length, $L_{\text{bh}}$, and the other assumed ground condition parameters.

Given the fluid temperature schedule determined, the instantaneous COP of the GSHP system is computed using the following empirical relationship from \citep{kensa2014HowCOPVaries},
\begin{equation}
    \text{COP}(t) = 4.0279 +0.1319 \cdot T_{\text{fluid}}(t)
\end{equation}
The energy that is provided by the GSHP system to the building is then given by,
\begin{equation}
    E_{\text{GSHP}}(t) = \frac{E_g(t)}{1-1/\text{COP}(t)}
\end{equation}
If the GSHP system is unable to meet the building load in any time instance, the remaining unsatisfied load is provided by an auxiliary heat supply system with a COP of 1.
\begin{equation}
    E_{\text{load}}(t) = E_{\text{GSHP}}(t) + E_{\text{aux}}(t)
\end{equation}
The total electricity consumption of the combined heat supply system is therefore given by,
\begin{equation}
    e_{\text{total}} = \sum_t \left( \frac{E_{\text{GSHP}}(t)}{\text{COP}(t)} + E_{\text{aux}}(t) \right)
\end{equation}
The cost of electricity used by the residential building is taken to be 32.6 p/kWh.

\begin{figure}[h]
    \centering
    \includegraphics[width=0.6\linewidth]{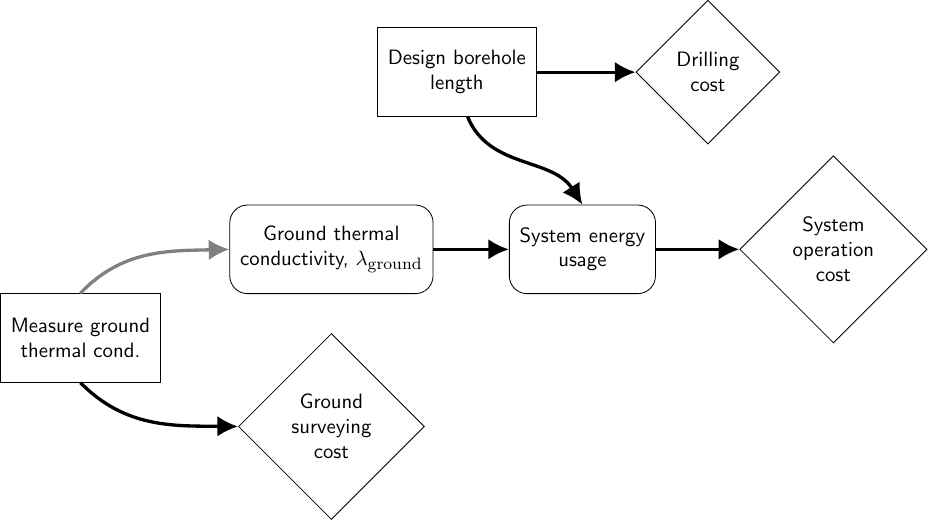}
    \vspace{4pt}
    \caption{Influence diagram representation of GSHP heat supply system design decision problem}
    \label{fig:ID-GSHP-design}
\end{figure}

The stochastic decision problem of designing borehole lengths as to minimise the expected lifetime cost of the heating supply system is represented as an influence diagram in Fig. \ref{fig:ID-GSHP-design}.\\

Solving the Prior Decision Problem, the optimal borehole length is found to be 155 m, leading to an expected overall lifetime cost of £819,100 for the heating system.

Various technologies exist for measuring ground thermal properties. Table \ref{tab:ground-tests} presents the cost and experimental uncertainty of thermal probe tests and Thermal Response Tests (TRT),  which are commonly used for ground thermal conductivity measurement. The extended TRT measurement is a hypothesised base case for the results achievable with TRT methods. The likelihood of obtaining a measurement $z$ for each test given a true underlying ground thermal conductivity of $\lambda$ is modelled using a Gaussian distribution as,
\begin{equation}
    f_e(z|\lambda) \sim \mathcal{N} \left( z,0,\frac{\nu_e}{2} \lambda \right)
\end{equation}
where $\nu_e$ is the decimal experimental uncertainty of the test.

\begin{table}[h]
    \centering
    \renewcommand{\arraystretch}{0.7}
    \begin{tabular}{c|c|c|c} \toprule \toprule
        Method & Uncertainty & Cost (£) & References \\ \midrule
        Thermal probe (in situ) & 25\% & 187 & \cite{king2012FieldDeterminationShallow,sunbeltrentals2023ThermtestTLS100Thermal} \\
        Thermal probe (lab test) & 17\% & 1,800 & \cite{low2013MeasuringSoilThermal,basaltridgetestinglaboratory2024BasaltRidgeTesting} \\
        Thermal Response Test (TRT) & 10\% & 5,000 & \cite{choi2021DevelopmentChillerattachedApparatus,spitler2000SituMeasurementGround,tang2019SensitiveAnalysisEffective} \\
        Extended TRT & 5\% & 10,000 & -- \\
        \bottomrule \bottomrule
    \end{tabular}
    \caption{Uncertainty and cost of ground thermal conductivity measurement tests}
    \label{tab:ground-tests}
\end{table}

\begin{figure}[h]
    \vspace*{-0.5cm}
    \centering
    \includegraphics[width=0.6\linewidth]{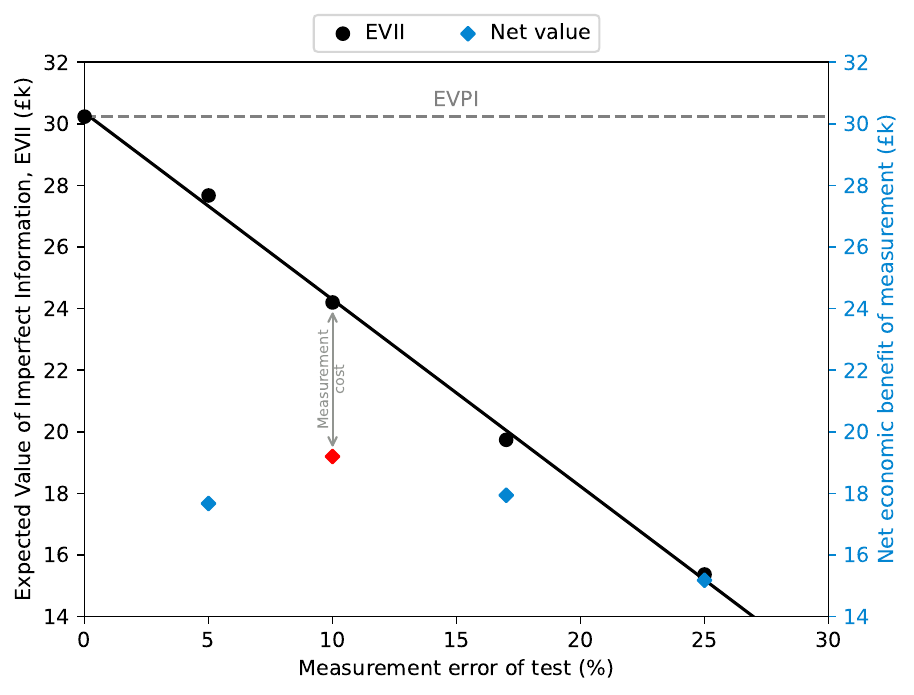}
    \vspace*{-0.2cm}
    \caption{Expected Value of Imperfect Information (EVII) and net economic benefit (EVII minus measurement cost) of imprecise ground thermal conductivity measurements}
    \label{fig:GSHP-EVIIs}
\end{figure}

The Expected Value of Imperfect Information (EVII) is computed for each available test, which is compared to the associated cost to determine the net benefit to decision making provided. Fig. \ref{fig:GSHP-EVIIs} plots the results. All tests provide significant net benefit to the heating system design task, indicating that site specific ground thermal testing should be conducted prior to the design of GSHP based heating systems. The TRT provides the greatest net added value to the decision maker, and so using this VoI analysis, the designer of this energy system can justify and optimise their expenditure on ground testing to support their decision making. This demonstrates that additional uncertainty reduction through more data collection or more precise measurement does not always provide sufficient improvements to decision making to warrant its cost.

\section{Conclusions} \label{sec:conclusions}

This study demonstrated the capabilities of Value of Information analysis (VoI) for providing a methodology to quantify the benefit of data collection to support decision making in building energy systems, and its use as a decision support tool for evaluating data collection strategies. Three example decision problems covering the management, operation, and design of building energy systems were studied: heat pump maintenance scheduling, office ventilation scheduling, and ground source heat pump system design.

Installing smart meters in ASHPs for dynamic maintenance scheduling was found to be a worthwhile investment. However, improvements in decision making provided were extremely limited, with perfect knowledge of heat pump condition leading to only a 0.06\% reduction in operational costs on average.
Occupancy monitoring systems enabled significant cost savings when scheduling ventilation to manage indoor air quality and infection risk in office spaces, reducing the total cost to the tenant by 10.2-16.4\%. Occupancy information was found to be more valuable in mid-sized offices and when infection is more prevalent, e.g. in the winter.
When designing a GSHP based heating system for a block of flats, ground thermal tests to reduce uncertainty in thermal conductivity were found to be beneficial to support system design. A Thermal Response Test (TRT) was determined to be the optimal data collection strategy, enabling net savings of £19k on the lifetime cost of the heating system.

In each example problem a different capability of the VoI framework was demonstrated: the ability to simply and clearly represent and analyse decisions in complex engineering systems, the identification of system characteristics for which data collection is worthwhile, and the determination of optimal data collection strategies when trading-off measurement precision versus cost.

However, applying VoI to practical energy systems presents several challenges. For instance, exact costs for equipment, installation, and maintenance are rarely known at the time of planning. Uncertainties in these costs, which determine the overall utility, should be included in the decision problem formulation and VoI analysis, increasing complexity. Additionally, it is often difficult to identify and model all of the decisions a proposed measurement can support. As a result, the VoI values calculated may miss significant contributions to the true value of data, and lead to under-collection of data. Further, at present applications of VoI assume decision makers to be risk neutral, and do not account for the role of data collection in managing risk. Within energy systems, risk management plays a significant role in planning, for example the over-sizing of heating systems to guarantee heating capacity and occupant thermal comfort. The inclusion of risk within decision objectives is necessary to properly value and prioritise data collection, but again increases analysis complexity.

{\color{\hlcolor}To enable the study of the economic benefit of data collection within buildings using VoI, further research is required to identify the practical decisions which could be supported by additional data collection, and develop the modelling infrastructure required to test its value. While detailed physics modelling of building energy systems is now widespread, more efficient simulation tools are required to enable the large number of sample simulations needed to estimate the statistical VoI metrics. Additionally, more detailed statistical modelling of key uncertainties within buildings\footnote{For instance, distributions of capital, installation, and operation costs for energy equipment, rather than just point estimates.} is required to achieve accurate VoI estimates, and avoid the need for costly, large-scale sensitivity analyses. The development of an ecosystem of key decision problems faced by practitioners, openly available modelling blocks for simulating (parts of) building energy systems, and statistical models of standard buildings cases (including assumptions and uncertainties), would allow researchers to systematically determine optimal data collection strategies. Additionally, with this framework, the economic merit of novel data collection methodologies could be evaluated without the need for expensive physical trials. As the quantity of data collected within buildings increases, the following questions become increasingly pertinent, ``What data provides value for decision making?'', ``What data is \textit{worth} collecting?''.}

Whilst the study of practical energy systems is challenging, VoI provides a clear methodology for addressing the present research gap of quantifying the benefit of data collection to support decision making in the building energy systems literature, and rationalising data collection strategies to avoid resource wastage on low insight data as building digitisation progresses. The insights into the benefits of data collection demonstrated in this work motivate {\color{\hlcolor}the} further study of VoI in the contexts of managing, operating, and designing building energy systems.
\newpage

\section*{CRediT authorship contribution statement}

\textbf{Max Langtry}: Conceptualization, Software, Methodology, Investigation, Writing - Original Draft, Project administration
\textbf{Chaoqun Zhuang}: Methodology
\textbf{Rebecca Ward}: Methodology
\textbf{Nikolas Makasis}: Methodology, Software
\textbf{Monika J. Kreitmair}: Methodology
\textbf{Zack Xuereb Conti}: Methodology
\textbf{Domenic Di Francesco}: Conceptualization, Supervision
\textbf{Ruchi Choudhary}: Conceptualization, Supervision, Writing - Review \& Editing

\section*{Declaration of competing interests}

The authors declare that they have no known competing financial interests or personal relationships that could have appeared to influence the work reported in this paper.

\section*{Data availability}

All code and data used to perform the experiments in this study is available 
at \url{https://github.com/EECi/VOI-for-Building-Energy}.

\section*{Acknowledgements}

Max Langtry is supported by the Engineering and Physical Sciences Research Council, through the CDT in Future Infrastructure and Built Environment: Resilience in a Changing World, Grant [EP/S02302X/1].

Chaoqun Zhuang, Rebecca Ward, Zack Xuereb Conti, and Domenic Di Francesco are supported by the Ecosystem Leadership Award under the EPSRC Grant [EP/X03870X/1], and The Alan Turing Institute, Grant [EP/N510129/1], particularly the Turing Research Fellowship scheme under that grant.

Nikolas Makasis and Monika J. Kreitmair are supported by the Surrey Future Fellowship program at the University of Surrey and CMMI-EPSRC: Modeling and Monitoring of Urban Underground Climate Change, Grant [EP/T019425/1].


\newpage
\bibliographystyle{elsarticle-num} 
\bibliography{VoI_refs}

\begin{thebibliography}{10}
\expandafter\ifx\csname url\endcsname\relax
  \def\url#1{\texttt{#1}}\fi
\expandafter\ifx\csname urlprefix\endcsname\relax\def\urlprefix{URL }\fi
\expandafter\ifx\csname href\endcsname\relax
  \def\href#1#2{#2} \def\path#1{#1}\fi

\bibitem{ccc2020SixthCarbonBudget}
CCC, \href{https://www.theccc.org.uk/wp-content/uploads/2020/12/Sector-summary-Buildings.pdf}{The {{Sixth Carbon Budget}} - {{Buildings}}}, Tech. rep., Committee on Climate Change (Dec. 2020).
\newline\urlprefix\url{https://www.theccc.org.uk/wp-content/uploads/2020/12/Sector-summary-Buildings.pdf}

\bibitem{tian2018ReviewUncertaintyAnalysis}
W.~Tian, Y.~Heo, P.~Wilde, Z.~Li, D.~Yan, C.~S. Park, X.~Feng, G.~Augenbroe, \href{https://www.sciencedirect.com/science/article/pii/S136403211830368X}{A review of uncertainty analysis in building energy assessment}, Renewable and Sustainable Energy Reviews 93 (2018) 285--301.
\newblock \href {https://doi.org/10.1016/j.rser.2018.05.029} {\path{doi:10.1016/j.rser.2018.05.029}}.
\newline\urlprefix\url{https://www.sciencedirect.com/science/article/pii/S136403211830368X}

\bibitem{yue2018ReviewApproachesUncertainty}
X.~Yue, S.~Pye, J.~DeCarolis, F.~G.~N. Li, F.~Rogan, B.~{\'O}. Gallach{\'o}ir, \href{https://www.sciencedirect.com/science/article/pii/S2211467X18300543}{A review of approaches to uncertainty assessment in energy system optimization models}, Energy Strategy Reviews 21 (2018) 204--217.
\newblock \href {https://doi.org/10.1016/j.esr.2018.06.003} {\path{doi:10.1016/j.esr.2018.06.003}}.
\newline\urlprefix\url{https://www.sciencedirect.com/science/article/pii/S2211467X18300543}

\bibitem{molina-solana2017DataScienceBuilding}
M.~{Molina-Solana}, M.~Ros, M.~D. Ruiz, J.~{G{\'o}mez-Romero}, M.~J. {Martin-Bautista}, \href{https://www.sciencedirect.com/science/article/pii/S1364032116308814}{Data science for building energy management: {{A}} review}, Renewable and Sustainable Energy Reviews 70 (2017) 598--609.
\newblock \href {https://doi.org/10.1016/j.rser.2016.11.132} {\path{doi:10.1016/j.rser.2016.11.132}}.
\newline\urlprefix\url{https://www.sciencedirect.com/science/article/pii/S1364032116308814}

\bibitem{rysanek2013OptimumBuildingEnergy}
A.~M. Rysanek, R.~Choudhary, \href{https://www.sciencedirect.com/science/article/pii/S0378778812005361}{Optimum building energy retrofits under technical and economic uncertainty}, Energy and Buildings 57 (2013) 324--337.
\newblock \href {https://doi.org/10.1016/j.enbuild.2012.10.027} {\path{doi:10.1016/j.enbuild.2012.10.027}}.
\newline\urlprefix\url{https://www.sciencedirect.com/science/article/pii/S0378778812005361}

\bibitem{mavromatidis2018ReviewUncertaintyCharacterisation}
G.~Mavromatidis, K.~Orehounig, J.~Carmeliet, \href{https://www.sciencedirect.com/science/article/pii/S1364032118300510}{A review of uncertainty characterisation approaches for the optimal design of distributed energy systems}, Renewable and Sustainable Energy Reviews 88 (2018) 258--277.
\newblock \href {https://doi.org/10.1016/j.rser.2018.02.021} {\path{doi:10.1016/j.rser.2018.02.021}}.
\newline\urlprefix\url{https://www.sciencedirect.com/science/article/pii/S1364032118300510}

\bibitem{mobaraki2022NovelDataAcquisition}
B.~Mobaraki, S.~Komarizadehasl, F.~J. Castilla~Pascual, J.~A. {Lozano-Galant}, R.~Porras~Soriano, \href{https://www.mdpi.com/2075-5309/12/5/670}{A {{Novel Data Acquisition System}} for {{Obtaining Thermal Parameters}} of {{Building Envelopes}}}, Buildings 12~(5) (2022) 670.
\newblock \href {https://doi.org/10.3390/buildings12050670} {\path{doi:10.3390/buildings12050670}}.
\newline\urlprefix\url{https://www.mdpi.com/2075-5309/12/5/670}

\bibitem{mantha2015RealTimeBuildingEnergy}
B.~Mantha, C.~Feng, C.~Menassa, V.~Kamat, \href{https://www.iaarc.org/publications/2015_proceedings_of_the_32st_isarc_oulu_finland/real_time_building_energy_and_comfort_parameter_data_collection_using_mobile_indoor_robots.html}{Real-{{Time Building Energy}} and {{Comfort Parameter Data Collection Using Mobile Indoor Robots}}}, ISARC Proceedings (2015) 1--9.
\newline\urlprefix\url{https://www.iaarc.org/publications/2015_proceedings_of_the_32st_isarc_oulu_finland/real_time_building_energy_and_comfort_parameter_data_collection_using_mobile_indoor_robots.html}

\bibitem{xia2014ComparisonBuildingEnergy}
J.~Xia, T.~Hong, Q.~Shen, W.~Feng, L.~Yang, P.~Im, A.~Lu, M.~Bhandari, \href{https://www.sciencedirect.com/science/article/pii/S0378778814003533}{Comparison of building energy use data between the {{United States}} and {{China}}}, Energy and Buildings 78 (2014) 165--175.
\newblock \href {https://doi.org/10.1016/j.enbuild.2014.04.031} {\path{doi:10.1016/j.enbuild.2014.04.031}}.
\newline\urlprefix\url{https://www.sciencedirect.com/science/article/pii/S0378778814003533}

\bibitem{han2020EnergysavingBuildingSystem}
K.~Han, J.~Zhang, \href{https://www.sciencedirect.com/science/article/pii/S0378778819329573}{Energy-saving building system integration with a smart and low-cost sensing/control network for sustainable and healthy living environments: {{Demonstration}} case study}, Energy and Buildings 214 (2020) 109861.
\newblock \href {https://doi.org/10.1016/j.enbuild.2020.109861} {\path{doi:10.1016/j.enbuild.2020.109861}}.
\newline\urlprefix\url{https://www.sciencedirect.com/science/article/pii/S0378778819329573}

\bibitem{luo2022ThreeyearDatasetSupporting}
N.~Luo, Z.~Wang, D.~Blum, C.~Weyandt, N.~Bourassa, M.~A. Piette, T.~Hong, \href{https://www.nature.com/articles/s41597-022-01257-x}{A three-year dataset supporting research on building energy management and occupancy analytics}, Scientific Data 9~(1) (2022) 156.
\newblock \href {https://doi.org/10.1038/s41597-022-01257-x} {\path{doi:10.1038/s41597-022-01257-x}}.
\newline\urlprefix\url{https://www.nature.com/articles/s41597-022-01257-x}

\bibitem{streltsov2020EstimatingResidentialBuilding}
A.~Streltsov, J.~M. Malof, B.~Huang, K.~Bradbury, \href{https://www.sciencedirect.com/science/article/pii/S0306261920314616}{Estimating residential building energy consumption using overhead imagery}, Applied Energy 280 (2020) 116018.
\newblock \href {https://doi.org/10.1016/j.apenergy.2020.116018} {\path{doi:10.1016/j.apenergy.2020.116018}}.
\newline\urlprefix\url{https://www.sciencedirect.com/science/article/pii/S0306261920314616}

\bibitem{johnzhai2020AssessingImplicationsSubmetering}
Z.~(John)~Zhai, A.~Salazar, \href{https://www.sciencedirect.com/science/article/pii/S2666123319300029}{Assessing the implications of submetering with energy analytics to building energy savings}, Energy and Built Environment 1~(1) (2020) 27--35.
\newblock \href {https://doi.org/10.1016/j.enbenv.2019.08.002} {\path{doi:10.1016/j.enbenv.2019.08.002}}.
\newline\urlprefix\url{https://www.sciencedirect.com/science/article/pii/S2666123319300029}

\bibitem{cho2019EnergyPerformanceAssessment}
K.~H. Cho, S.~S. Kim, \href{https://www.mdpi.com/1996-1073/12/6/1149}{Energy {{Performance Assessment According}} to {{Data Acquisition Levels}} of {{Existing Buildings}}}, Energies 12~(6) (2019) 1149.
\newblock \href {https://doi.org/10.3390/en12061149} {\path{doi:10.3390/en12061149}}.
\newline\urlprefix\url{https://www.mdpi.com/1996-1073/12/6/1149}

\bibitem{winschermann2023AssessingValueInformation}
L.~Winschermann, N.~Ba{\~n}ol~Arias, G.~Hoogsteen, J.~Hurink, \href{https://www.sciencedirect.com/science/article/pii/S1364032123004574}{Assessing the value of information for electric vehicle charging strategies at office buildings}, Renewable and Sustainable Energy Reviews 185 (2023) 113600.
\newblock \href {https://doi.org/10.1016/j.rser.2023.113600} {\path{doi:10.1016/j.rser.2023.113600}}.
\newline\urlprefix\url{https://www.sciencedirect.com/science/article/pii/S1364032123004574}

\bibitem{langtry2024ImpactDataForecasting}
M.~Langtry, V.~Wichitwechkarn, R.~Ward, C.~Zhuang, M.~J. Kreitmair, N.~Makasis, Z.~Xuereb~Conti, R.~Choudhary, \href{https://www.sciencedirect.com/science/article/pii/S0378778824007217}{Impact of data for forecasting on performance of model predictive control in buildings with smart energy storage}, Energy and Buildings 320 (2024) 114605.
\newblock \href {https://doi.org/10.1016/j.enbuild.2024.114605} {\path{doi:10.1016/j.enbuild.2024.114605}}.
\newline\urlprefix\url{https://www.sciencedirect.com/science/article/pii/S0378778824007217}

\bibitem{glasgo2017AssessingValueInformation}
B.~Glasgo, C.~Hendrickson, I.~L. Azevedo, \href{https://www.sciencedirect.com/science/article/pii/S0306261917307237}{Assessing the value of information in residential building simulation: {{Comparing}} simulated and actual building loads at the circuit level}, Applied Energy 203 (2017) 348--363.
\newblock \href {https://doi.org/10.1016/j.apenergy.2017.05.164} {\path{doi:10.1016/j.apenergy.2017.05.164}}.
\newline\urlprefix\url{https://www.sciencedirect.com/science/article/pii/S0306261917307237}

\bibitem{risch2021InfluenceDataAcquisition}
S.~Risch, P.~Remmen, D.~M{\"u}ller, \href{https://www.sciencedirect.com/science/article/pii/S0378778820313748}{Influence of data acquisition on the {{Bayesian}} calibration of urban building energy models}, Energy and Buildings 230 (2021) 110512.
\newblock \href {https://doi.org/10.1016/j.enbuild.2020.110512} {\path{doi:10.1016/j.enbuild.2020.110512}}.
\newline\urlprefix\url{https://www.sciencedirect.com/science/article/pii/S0378778820313748}

\bibitem{tian2016IdentifyingInformativeEnergy}
W.~Tian, S.~Yang, Z.~Li, S.~Wei, W.~Pan, Y.~Liu, \href{https://www.sciencedirect.com/science/article/pii/S0378778816301967}{Identifying informative energy data in {{Bayesian}} calibration of building energy models}, Energy and Buildings 119 (2016) 363--376.
\newblock \href {https://doi.org/10.1016/j.enbuild.2016.03.042} {\path{doi:10.1016/j.enbuild.2016.03.042}}.
\newline\urlprefix\url{https://www.sciencedirect.com/science/article/pii/S0378778816301967}

\bibitem{wang2022DataAcquisitionUrban}
C.~Wang, M.~Ferrando, F.~Causone, X.~Jin, X.~Zhou, X.~Shi, \href{https://www.sciencedirect.com/science/article/pii/S0360132322002955}{Data acquisition for urban building energy modeling: {{A}} review}, Building and Environment 217 (2022) 109056.
\newblock \href {https://doi.org/10.1016/j.buildenv.2022.109056} {\path{doi:10.1016/j.buildenv.2022.109056}}.
\newline\urlprefix\url{https://www.sciencedirect.com/science/article/pii/S0360132322002955}

\bibitem{han2021ApproachDataAcquisition}
M.~Han, Z.~Wang, X.~Zhang, \href{https://www.mdpi.com/2075-5309/11/1/30}{An {{Approach}} to {{Data Acquisition}} for {{Urban Building Energy Modeling Using}} a {{Gaussian Mixture Model}} and {{Expectation-Maximization Algorithm}}}, Buildings 11~(1) (2021) 30.
\newblock \href {https://doi.org/10.3390/buildings11010030} {\path{doi:10.3390/buildings11010030}}.
\newline\urlprefix\url{https://www.mdpi.com/2075-5309/11/1/30}

\bibitem{raiffa1969ReviewDecisionAnalysis}
H.~Raiffa, \href{https://www.jstor.org/stable/2325714}{Review of {{Decision Analysis}}: {{Introductory Lectures}} on {{Choices Under Uncertainty}}.}, The Journal of Finance 24~(5) (1969) 1000--1003.
\newblock \href {http://arxiv.org/abs/2325714} {\path{arXiv:2325714}}, \href {https://doi.org/10.2307/2325714} {\path{doi:10.2307/2325714}}.
\newline\urlprefix\url{https://www.jstor.org/stable/2325714}

\bibitem{howard1966InformationValueTheory}
R.~A. Howard, \href{https://ieeexplore.ieee.org/abstract/document/4082064}{Information {{Value Theory}}}, IEEE Transactions on Systems Science and Cybernetics 2~(1) (1966) 22--26.
\newblock \href {https://doi.org/10.1109/TSSC.1966.300074} {\path{doi:10.1109/TSSC.1966.300074}}.
\newline\urlprefix\url{https://ieeexplore.ieee.org/abstract/document/4082064}

\bibitem{zhang2021ValueInformationAnalysis}
W.-H. Zhang, D.-G. Lu, J.~Qin, S.~Th{\"o}ns, M.~H. Faber, \href{https://doi.org/10.1186/s43065-021-00027-0}{Value of information analysis in civil and infrastructure engineering: A review}, Journal of Infrastructure Preservation and Resilience 2~(1) (2021) 16.
\newblock \href {https://doi.org/10.1186/s43065-021-00027-0} {\path{doi:10.1186/s43065-021-00027-0}}.
\newline\urlprefix\url{https://doi.org/10.1186/s43065-021-00027-0}

\bibitem{pratt1995IntroductionStatisticalDecision}
J.~W. Pratt, H.~Raiffa, R.~Schlaifer, Introduction to {{Statistical Decision Theory}}, MIT Press, 1995.

\bibitem{keisler2014ValueInformationAnalysis}
J.~M. Keisler, Z.~A. Collier, E.~Chu, N.~Sinatra, I.~Linkov, \href{https://doi.org/10.1007/s10669-013-9439-4}{Value of information analysis: The state of application}, Environment Systems and Decisions 34~(1) (2014) 3--23.
\newblock \href {https://doi.org/10.1007/s10669-013-9439-4} {\path{doi:10.1007/s10669-013-9439-4}}.
\newline\urlprefix\url{https://doi.org/10.1007/s10669-013-9439-4}

\bibitem{grussing2018OptimizedBuildingComponent}
M.~N. Grussing, \href{https://ascelibrary.org/doi/10.1061/%28ASCE%29CF.1943-5509.0001198}{Optimized {{Building Component Assessment Planning Using}} a {{Value}} of {{Information Model}}}, Journal of Performance of Constructed Facilities 32~(4) (2018) 04018054.
\newblock \href {https://doi.org/10.1061/(ASCE)CF.1943-5509.0001198} {\path{doi:10.1061/(ASCE)CF.1943-5509.0001198}}.
\newline\urlprefix\url{https://ascelibrary.org/doi/10.1061/%28ASCE%29CF.1943-5509.0001198}

\bibitem{myklebust2020ValueInformationAnalysis}
H.~O.~V. Myklebust, J.~Eidsvik, I.~B. Sperstad, D.~Bhattacharjya, \href{https://pubsonline.informs.org/doi/abs/10.1287/deca.2019.0405}{Value of {{Information Analysis}} for {{Complex Simulator Models}}: {{Application}} to {{Wind Farm Maintenance}}}, Decision Analysis 17~(2) (2020) 134--153.
\newblock \href {https://doi.org/10.1287/deca.2019.0405} {\path{doi:10.1287/deca.2019.0405}}.
\newline\urlprefix\url{https://pubsonline.informs.org/doi/abs/10.1287/deca.2019.0405}

\bibitem{esnaasharyesfahani2020PrioritizingPreprojectPlanning}
M.~Esnaashary~Esfahani, C.~Rausch, C.~Haas, B.~T. Adey, \href{https://ascelibrary.org/doi/10.1061/%28ASCE%29ME.1943-5479.0000822}{Prioritizing {{Preproject Planning Activities Using Value}} of {{Information Analysis}}}, Journal of Management in Engineering 36~(5) (2020) 04020064.
\newblock \href {https://doi.org/10.1061/(ASCE)ME.1943-5479.0000822} {\path{doi:10.1061/(ASCE)ME.1943-5479.0000822}}.
\newline\urlprefix\url{https://ascelibrary.org/doi/10.1061/%28ASCE%29ME.1943-5479.0000822}

\bibitem{malings2016ValueInformationSpatially}
C.~Malings, M.~Pozzi, \href{https://www.sciencedirect.com/science/article/pii/S0951832016300564}{Value of information for spatially distributed systems: {{Application}} to sensor placement}, Reliability Engineering \& System Safety 154 (2016) 219--233.
\newblock \href {https://doi.org/10.1016/j.ress.2016.05.010} {\path{doi:10.1016/j.ress.2016.05.010}}.
\newline\urlprefix\url{https://www.sciencedirect.com/science/article/pii/S0951832016300564}

\bibitem{difrancesco2021DecisiontheoreticInspectionPlanning}
D.~Di~Francesco, M.~Chryssanthopoulos, M.~H. Faber, U.~Bharadwaj, \href{https://www.cambridge.org/core/product/identifier/S2632673621000186/type/journal_article}{Decision-theoretic inspection planning using imperfect and incomplete data}, Data-Centric Engineering 2 (2021) e18.
\newblock \href {https://doi.org/10.1017/dce.2021.18} {\path{doi:10.1017/dce.2021.18}}.
\newline\urlprefix\url{https://www.cambridge.org/core/product/identifier/S2632673621000186/type/journal_article}

\bibitem{difrancesco2023SystemEffectsIdentifying}
D.~Di~Francesco, M.~Langtry, A.~B. Duncan, C.~Dent, \href{http://arxiv.org/abs/2309.07695}{System {{Effects}} in {{Identifying Risk-Optimal Data Requirements}} for {{Digital Twins}} of {{Structures}}} (Sep. 2023).
\newblock \href {http://arxiv.org/abs/2309.07695} {\path{arXiv:2309.07695}}, \href {https://doi.org/10.48550/arXiv.2309.07695} {\path{doi:10.48550/arXiv.2309.07695}}.
\newline\urlprefix\url{http://arxiv.org/abs/2309.07695}

\bibitem{niu2023FrameworkQuantifyingValue}
J.~Niu, X.~Li, Z.~Tian, H.~Yang, \href{https://www.sciencedirect.com/science/article/pii/S0306261923010814}{A framework for quantifying the value of information to mitigate risk in the optimal design of distributed energy systems under uncertainty}, Applied Energy 350 (2023) 121717.
\newblock \href {https://doi.org/10.1016/j.apenergy.2023.121717} {\path{doi:10.1016/j.apenergy.2023.121717}}.
\newline\urlprefix\url{https://www.sciencedirect.com/science/article/pii/S0306261923010814}

\bibitem{smith1945TheoryGamesEconomic}
C.~a.~B. Smith, \href{https://www.cambridge.org/core/journals/mathematical-gazette/article/abs/theory-of-games-and-economic-behaviour-by-john-von-neumann-and-oskar-morgenstern-pp-xviii-625-66s-6d-1944-princeton-up-humphrey-milford/76E26E70D0D9B6AF26DB67A6A10D58CF}{Theory of {{Games}} and {{Economic Behaviour}}. {{By John Von Neumann}} and {{Oskar Morgenstern}}. {{Pp}}. xviii, 625. 66s. 6d. 1944. ({{Princeton U}}.{{P}}.; {{Humphrey Milford}})}, The Mathematical Gazette 29~(285) (1945) 131--133.
\newblock \href {https://doi.org/10.2307/3610940} {\path{doi:10.2307/3610940}}.
\newline\urlprefix\url{https://www.cambridge.org/core/journals/mathematical-gazette/article/abs/theory-of-games-and-economic-behaviour-by-john-von-neumann-and-oskar-morgenstern-pp-xviii-625-66s-6d-1944-princeton-up-humphrey-milford/76E26E70D0D9B6AF26DB67A6A10D58CF}

\bibitem{difrancesco2023GuidanceUseProbabilistic}
D.~Di~Francesco, \href{https://github.com/DomDF/DCE_guidance}{Guidance on the {{Use}} of {{Probabilistic Methods}} for {{Identifying Data Requirements}}} (Feb. 2023).
\newline\urlprefix\url{https://github.com/DomDF/DCE_guidance}

\bibitem{pickering2021QuantifyingResilienceEnergy}
B.~Pickering, R.~Choudhary, \href{https://www.sciencedirect.com/science/article/pii/S0306261921000313}{Quantifying resilience in energy systems with out-of-sample testing}, Applied Energy 285 (2021) 116465.
\newblock \href {https://doi.org/10.1016/j.apenergy.2021.116465} {\path{doi:10.1016/j.apenergy.2021.116465}}.
\newline\urlprefix\url{https://www.sciencedirect.com/science/article/pii/S0306261921000313}

\bibitem{birge1982ValueStochasticSolution}
J.~R. Birge, \href{https://doi.org/10.1007/BF01585113}{The value of the stochastic solution in stochastic linear programs with fixed recourse}, Mathematical Programming 24~(1) (1982) 314--325.
\newblock \href {https://doi.org/10.1007/BF01585113} {\path{doi:10.1007/BF01585113}}.
\newline\urlprefix\url{https://doi.org/10.1007/BF01585113}

\bibitem{departmentforenergysecurityandnetzero2023AnnualDomesticEnergy}
{Department for Energy Security {and} Net Zero}, \href{https://www.gov.uk/government/statistical-data-sets/annual-domestic-energy-price-statistics}{Annual domestic energy bills} (Dec. 2023).
\newline\urlprefix\url{https://www.gov.uk/government/statistical-data-sets/annual-domestic-energy-price-statistics}

\bibitem{nouvel2015EuropeanMappingSeasonal}
R.~Nouvel, M.~Cotrado, D.~Pietruschka, European mapping of seasonal performances of air-source and geothermal heat pumps for residential applications, Proceedings of International Conference CISBAT 2015 Future Buildings and Districts Sustainability from Nano to Urban Scale (2015).

\bibitem{griffith2008MethodologyModelingBuilding}
B.~Griffith, N.~Long, P.~Torcellini, R.~Judkoff, D.~Crawley, J.~Ryan, \href{https://www.osti.gov/biblio/926505}{Methodology for {{Modeling Building Energy Performance}} across the {{Commercial Sector}}}, Tech. Rep. NREL/TP-550-41956, National Renewable Energy Lab. (NREL), Golden, CO (United States) (Mar. 2008).
\newblock \href {https://doi.org/10.2172/926505} {\path{doi:10.2172/926505}}.
\newline\urlprefix\url{https://www.osti.gov/biblio/926505}

\bibitem{daikin2022DaikinUKPrice}
Daikin, \href{https://sbs.co.uk/file/Supplier/Daikin_Price_List.pdf}{Daikin {{UK}} price book}, Tech. rep. (Apr. 2022).
\newline\urlprefix\url{https://sbs.co.uk/file/Supplier/Daikin_Price_List.pdf}

\bibitem{deoliveira2021EvolutionSprayAerosol}
P.~M. {de Oliveira}, L.~C.~C. Mesquita, S.~Gkantonas, A.~Giusti, E.~Mastorakos, \href{https://royalsocietypublishing.org/doi/full/10.1098/rspa.2020.0584}{Evolution of spray and aerosol from respiratory releases: Theoretical estimates for insight on viral transmission}, Proceedings of the Royal Society A: Mathematical, Physical and Engineering Sciences 477~(2245) (2021) 20200584.
\newblock \href {https://doi.org/10.1098/rspa.2020.0584} {\path{doi:10.1098/rspa.2020.0584}}.
\newline\urlprefix\url{https://royalsocietypublishing.org/doi/full/10.1098/rspa.2020.0584}

\bibitem{gkantonas2021AirborneCamRisk}
S.~Gkantonas, D.~Zabotti, L.~C. Mesquita, E.~Mastorakos, P.~M. {de Oliveira}, \href{https://www.repository.cam.ac.uk/handle/1810/324994}{Airborne.cam: A risk calculator of {{SARS-CoV-2}} aerosol transmission under well-mixed ventilation conditions} (Jul. 2021).
\newline\urlprefix\url{https://www.repository.cam.ac.uk/handle/1810/324994}

\bibitem{ons2023CoronavirusCOVID19Latest}
ONS, \href{https://www.ons.gov.uk/peoplepopulationandcommunity/healthandsocialcare/conditionsanddiseases/articles/coronaviruscovid19latestinsights/infections}{Coronavirus ({{COVID-19}}) latest insights - {{Office}} for {{National Statistics}}} (Mar. 2023).
\newline\urlprefix\url{https://www.ons.gov.uk/peoplepopulationandcommunity/healthandsocialcare/conditionsanddiseases/articles/coronaviruscovid19latestinsights/infections}

\bibitem{ons2022EmployeeEarningsUK}
ONS, \href{https://www.ons.gov.uk/employmentandlabourmarket/peopleinwork/earningsandworkinghours/bulletins/annualsurveyofhoursandearnings/2022}{Employee earnings in the {{UK}} - {{Office}} for {{National Statistics}}} (Oct. 2022).
\newline\urlprefix\url{https://www.ons.gov.uk/employmentandlabourmarket/peopleinwork/earningsandworkinghours/bulletins/annualsurveyofhoursandearnings/2022}

\bibitem{aresti2018ReviewDesignAspects}
L.~Aresti, P.~Christodoulides, G.~Florides, \href{https://www.sciencedirect.com/science/article/pii/S1364032118302727}{A review of the design aspects of ground heat exchangers}, Renewable and Sustainable Energy Reviews 92 (2018) 757--773.
\newblock \href {https://doi.org/10.1016/j.rser.2018.04.053} {\path{doi:10.1016/j.rser.2018.04.053}}.
\newline\urlprefix\url{https://www.sciencedirect.com/science/article/pii/S1364032118302727}

\bibitem{dallasanta2020UpdatedGroundThermal}
G.~Dalla~Santa, A.~Galgaro, R.~Sassi, M.~Cultrera, P.~Scotton, J.~Mueller, D.~Bertermann, D.~Mendrinos, R.~Pasquali, R.~Perego, S.~Pera, E.~Di~Sipio, G.~Cassiani, M.~De~Carli, A.~Bernardi, \href{https://www.sciencedirect.com/science/article/pii/S0375650519301944}{An updated ground thermal properties database for {{GSHP}} applications}, Geothermics 85 (2020) 101758.
\newblock \href {https://doi.org/10.1016/j.geothermics.2019.101758} {\path{doi:10.1016/j.geothermics.2019.101758}}.
\newline\urlprefix\url{https://www.sciencedirect.com/science/article/pii/S0375650519301944}

\bibitem{busby2018ModellingStudyVariation}
J.~Busby, \href{https://doi.org/10.1144/qjegh2017-127}{A modelling study of the variation of thermal conductivity of the {{English Chalk}}}, Quarterly Journal of Engineering Geology and Hydrogeology 51~(4) (2018) 417--423.
\newblock \href {https://doi.org/10.1144/qjegh2017-127} {\path{doi:10.1144/qjegh2017-127}}.
\newline\urlprefix\url{https://doi.org/10.1144/qjegh2017-127}

\bibitem{busby2011ProvisionThermalProperties}
J.~Busby, \href{https://mail.gshp.org.uk/pdf/CambridgeSeminar2011/JonBusby_BGS_Thermal_Conductivity.pdf}{Provision of thermal properties data for ground collector loop design} (2011).
\newline\urlprefix\url{https://mail.gshp.org.uk/pdf/CambridgeSeminar2011/JonBusby_BGS_Thermal_Conductivity.pdf}

\bibitem{loveridge2013ThermalResponseTesting}
F.~Loveridge, G.~Holmes, W.~Powrie, T.~Roberts, \href{https://www.icevirtuallibrary.com/doi/10.1680/geng.12.00037}{Thermal response testing through the {{Chalk}} aquifer in {{London}}, {{UK}}}, Proceedings of the Institution of Civil Engineers - Geotechnical Engineering 166~(2) (2013) 197--210.
\newblock \href {https://doi.org/10.1680/geng.12.00037} {\path{doi:10.1680/geng.12.00037}}.
\newline\urlprefix\url{https://www.icevirtuallibrary.com/doi/10.1680/geng.12.00037}

\bibitem{mitchell2020UKPassivhausEnergy}
R.~Mitchell, S.~Natarajan, \href{https://www.sciencedirect.com/science/article/pii/S0378778820313918}{{{UK Passivhaus}} and the energy performance gap}, Energy and Buildings 224 (2020) 110240.
\newblock \href {https://doi.org/10.1016/j.enbuild.2020.110240} {\path{doi:10.1016/j.enbuild.2020.110240}}.
\newline\urlprefix\url{https://www.sciencedirect.com/science/article/pii/S0378778820313918}

\bibitem{lamarche2007NewContributionFinite}
L.~Lamarche, B.~Beauchamp, \href{https://www.sciencedirect.com/science/article/pii/S0378778806001824}{A new contribution to the finite line-source model for geothermal boreholes}, Energy and Buildings 39~(2) (2007) 188--198.
\newblock \href {https://doi.org/10.1016/j.enbuild.2006.06.003} {\path{doi:10.1016/j.enbuild.2006.06.003}}.
\newline\urlprefix\url{https://www.sciencedirect.com/science/article/pii/S0378778806001824}

\bibitem{kensa2014HowCOPVaries}
Kensa, \href{https://www.kensaheatpumps.com/wp-content/uploads/2014/03/Factsheet-COP-Variation-V2.pdf}{How {{COP}} varies with {{Inlet}} and {{Outlet Temperature V2}}}, Tech. rep. (Mar. 2014).
\newline\urlprefix\url{https://www.kensaheatpumps.com/wp-content/uploads/2014/03/Factsheet-COP-Variation-V2.pdf}

\bibitem{king2012FieldDeterminationShallow}
W.~King, D.~Banks, J.~Findlay, \href{https://doi.org/10.1144/qjegh2012-002}{Field determination of shallow soil thermal conductivity using a short-duration needle probe test}, Quarterly Journal of Engineering Geology and Hydrogeology 45~(4) (2012) 497--504.
\newblock \href {https://doi.org/10.1144/qjegh2012-002} {\path{doi:10.1144/qjegh2012-002}}.
\newline\urlprefix\url{https://doi.org/10.1144/qjegh2012-002}

\bibitem{sunbeltrentals2023ThermtestTLS100Thermal}
S.~Rentals, \href{https://web.archive.org/web/20231206070141/https://www.inlec.com/thermtest-tls-100}{Thermtest {{TLS-100 Thermal Conductivity Meter}}}, Tech. rep. (Dec. 2023).
\newline\urlprefix\url{https://web.archive.org/web/20231206070141/https://www.inlec.com/thermtest-tls-100}

\bibitem{low2013MeasuringSoilThermal}
J.~E. Low, F.~Loveridge, W.~Powrie, \href{https://www.cfms-sols.org/sites/default/files/Actes/3375-3378.pdf}{Measuring soil thermal properties for use in energy foundation design}, in: Proceedings of the 18th {{International Conference}} on {{Soil Mechanics}} and {{Geotechnical Engineering}}, {{Paris}} 2013, 2013, pp. 3375--3378.
\newline\urlprefix\url{https://www.cfms-sols.org/sites/default/files/Actes/3375-3378.pdf}

\bibitem{basaltridgetestinglaboratory2024BasaltRidgeTesting}
B.~R.~T. Laboratory, \href{https://web.archive.org/web/20240305171909/https://www.basaltridgetesting.com/pricing}{Basalt {{Ridge Testing Laboratory Price List}}}, Tech. rep. (Mar. 2024).
\newline\urlprefix\url{https://web.archive.org/web/20240305171909/https://www.basaltridgetesting.com/pricing}

\bibitem{choi2021DevelopmentChillerattachedApparatus}
W.~Choi, R.~Choudhary, R.~Ooka, \href{https://www.sciencedirect.com/science/article/pii/S0378778821001250}{Development of chiller-attached apparatus for accurate initial ground temperature measurement: {{Insights}} from global sensitivity analysis of thermal response tests}, Energy and Buildings 238 (2021) 110841.
\newblock \href {https://doi.org/10.1016/j.enbuild.2021.110841} {\path{doi:10.1016/j.enbuild.2021.110841}}.
\newline\urlprefix\url{https://www.sciencedirect.com/science/article/pii/S0378778821001250}

\bibitem{spitler2000SituMeasurementGround}
J.~D. Spitler, C.~Yavuzturk, S.~J. Rees, \href{https://citeseerx.ist.psu.edu/document?repid=rep1&type=pdf&doi=3856e0aadf680f5704e60f6a6c0aea32ac674e08}{In {{Situ Measurement}} of {{Ground Thermal Properties}}}, in: Proceedings of {{Terrastock}} 2000, 2000, pp. 165--170.
\newline\urlprefix\url{https://citeseerx.ist.psu.edu/document?repid=rep1&type=pdf&doi=3856e0aadf680f5704e60f6a6c0aea32ac674e08}

\bibitem{tang2019SensitiveAnalysisEffective}
F.~Tang, H.~Nowamooz, \href{https://www.sciencedirect.com/science/article/pii/S0960148119307931}{Sensitive analysis on the effective soil thermal conductivity of the {{Thermal Response Test}} considering various testing times, field conditions and {{U-pipe}} lengths}, Renewable Energy 143 (2019) 1732--1743.
\newblock \href {https://doi.org/10.1016/j.renene.2019.05.120} {\path{doi:10.1016/j.renene.2019.05.120}}.
\newline\urlprefix\url{https://www.sciencedirect.com/science/article/pii/S0960148119307931}

\end{thebibliography}

\end{document}